\def\ip{I_{peak}}
\def\tdur{t_{dur}}
\def\hvdk{{HK89}}

\newcommand{\LH}{2}

\newcommand{\SingleSpace}{
  \renewcommand{\LH}{0.90}
  \def\baselinestretch{\LH}
  \tiny
  \normalsize
}
\newlength{\parindnt}
\newcommand{\newpar}{\hspace{\parindnt}}
\newcommand{\PSbox}[3]{\mbox{\rule{0in}{#3}\includegraphics{#1}\hspace{#2}}}
\newcommand{\FigNum}[1]{\unitlength 1pt \begin{picture}(55,10)(-400,35) 
			\put(0,0){Figure #1}
			\end{picture}}

\newcommand{\Section}[1]{\section{#1} \newpar 
  \def\baselinestretch{\LH}
  \tiny
  \normalsize
}
\newcommand{\SectionN}[1]{\section*{#1} \newpar 
  \def\baselinestretch{\LH}
  \tiny
  \normalsize
}
\newcommand{\SubSection}[1]{\subsection{#1} \newpar 
  \def\baselinestretch{\LH}
  \tiny
  \normalsize
}

\newcommand{\SubSubSection}[1]{\subsubsection{#1} \newpar 
  \def\baselinestretch{\LH}
  \tiny
  \normalsize
}

\newcommand\sig{$\sigma$}

\newcommand\MIT{Massachusetts Institute of Technology, Room 37-627,
Cambridge, MA 02139, U.S.A.}
\newcommand\UVA{Astronomical Institute ``Anton Pannekoek'', University
of Amsterdam, Center for High Energy Astrophysics, Kruislaan 403, 1098
SJ Amsterdam, The Netherlands}

\newcommand\PUO{Princeton University Observatory, Peyton Hall, Princeton, NJ 08544-1001, U.S.A.}
\renewcommand\approx{\mbox{$\sim$}}
\newcommand\about{\mbox{$\sim$}}

\newcommand\approxgt{\mbox{$^{>}\hspace{-0.24cm}_{\sim}$}}
\newcommand\approxlt{\mbox{$^{<}\hspace{-0.24cm}_{\sim}$}}

\newcommand\ie{{\it i.e.}}
\newcommand\eg{{\it e.g.}}

\newcommand\etal{et~al.$\!$}

\def\Ref #1 {\lbrack {#1}\rbrack}
\def\vol#1  {{{\bf #1}{\rm,}\ }}
\def\etal   {{et~al.}\ }
\def\apj    {{ApJ{\rm,}\ }}
\def\apjl   {{ApJLett{\rm,}\ }}

\def\apjs   {{ApJS{\rm,}\ }}

\def\aa     {{A\&A{\rm,}\ }}

\def\mnras  {{MNRAS{\rm,}\ }}
\def\nat    {{Nat{\rm,}\ }}
\def\ssr    {{SSR{\rm,}\ }}
\def\pasj   {{PASJ{\rm,}\ }}

\documentstyle[12pt,aaspp]{article}
\begin{document}
\begin{center}
MNRAS, Submitted February 10, 1995; Accepted June 16, 1995 \\
\vskip 3cm
{\bf X-ray Timing and Spectral Behavior of the Rapid Burster}
\vskip 1.5cm
Robert E. Rutledge$^{1}$, Lori M. Lubin$^{2}$, Walter H.G.
Lewin$^{1}$, \\ Brian Vaughan$^3$, Jan van Paradijs$^{3, 4}$, and Michiel van der Klis$^{3}$
\vskip 2cm
\end{center}
\hspace*{-.25cm}
${^1}$ \MIT\\
\hspace*{-.25cm}
${^2}$ \PUO\\
\hspace*{-.25cm}
${^3}$ \UVA \\
\hspace*{-.25cm}
${^4}$ Physics Department, University of Alabama in Huntsville,
Huntsville AL 35899
\vskip 2cm
\pagebreak
\SingleSpace
\SectionN{Abstract}
We present an X-ray fast-timing and spectral analysis of the type II
bursts and the persistent emission (PE) of the Rapid Burster observed
with the {\it EXOSAT} Medium-Energy Instrument.  The
Hardness-Intensity and Color-Color Diagrams of the Rapid Burster are
somewhat different from those of the majority of other low-mass X-ray
binaries, which fall into two distinct groups, ``atoll''
and ``Z'' sources. The strength and frequencies of quasi-periodic
oscillations (QPO) in the bursts and the PE, as well as a strong
anti-correlation between QPO frequency and burst peak flux, also
distinguish the Rapid Burster from atoll and Z sources.

The presence, frequency, and strength of QPO in type II bursts are all
correlated in some manner to the spectral hardness, as are the
strength of QPO and the very low-frequency noise (VLFN) component in
the persistent emission.

\Section{Introduction}
Discovered in 1976 by Lewin \etal (1976), the Rapid Burster (MXB
1730-335, or ``RB'' herein) is located in the highly reddened globular
cluster Liller I (\cite{liller77}) at a distance of $\sim$ 10 kpc
(\cite{kleinmann76}).  The Rapid Burster is the only low-mass X-ray
binary (LMXB) which produces two types of X-ray bursts
(\cite{hoffman78}).  Type I bursts, which are observed from $\sim$ 40
other LMXB, are due to thermonuclear flashes on the surface of an
accreting neutron star (NS).  Type II bursts, which are unique to the
Rapid Burster, are the result of accretion instabilities
(gravitational potential energy).  The behavior of the type II bursts
can be described as that of a relaxation oscillator; the integrated
flux (E) in a burst is roughly proportional to the time interval
($\Delta$t) to the following burst (\cite{lewin76};
\cite{lewin77}; \cite{white78}; \cite{marshall79};
\cite{inoue80}).  For a detailed review, see Lewin, Van Paradijs \&
Taam (1993).

The Rapid Burster exhibits at least two type II burst modes: mode I is
characterized by bursts with a bimodal duration distribution, while
mode II has a unimodal duration distribution (\cite{marshall79}).  At
times type II bursts are observed with strong persistent X-ray
emission (PE) after long ($> 30$ sec) type II bursts.  This persistent
emission emerges gradually after a long type II burst and disappears
before the occurrence of the next type II burst; this disappearance
just before and after bursts is referred to as the ``dips'' in the
persistent emission (\cite{marshall79}; \cite{vpds79};
\cite{stella88a}). These dips are not likely to be due to an
obscuration of the source (\cite{lubin93}).  The persistent emission
between bursts also exhibits distinct features such as strong 40 mHz
oscillations, small bumps and glitches, and a characteristic ``hump''
in its profile (\cite{lubin92b}, 1993).

Lubin \etal (1992a) reported that the relation between burst peak
luminosity and peak black-body temperature depends only on the
characteristic mode of recurrence of the type II bursts, independent
of their duration (from $\sim$ 2 to $\sim$ 680 sec).  For bursts
during the mode II (single distribution of durations) there exists a
correlation between peak luminosity and black-body temperature
(Kunieda \etal 1984; Kawai \etal 1990; Tan \etal 1991) while bursts in
mode I (bimodal distribution of durations) have an approximately
constant temperature (\cite{lubin92a}). For type II bursts during mode
II, Kawai \etal (1990) and Tan \etal (1991) found an anti-correlation
between blackbody temperature and instantaneous burst flux.

Many type II bursts with durations in excess of $\sim$ 9 sec show
quasi-periodic oscillations (QPO) in the $\sim$ 2 - 7 Hz range
(\cite{tawara82}; \cite{stella88a}; \cite{dotani90}; \cite{lubin91}).
QPO have also been seen in bursts with durations as short as $\sim$ 3
sec (\cite{rutledge93}).  The average QPO centroid frequency during a
burst is strongly anti-correlated with the average burst peak flux
(except perhaps at low peak fluxes where they may be correlated; see
\cite{lubin91}).  QPO with frequencies of 2-5 Hz have also been discovered in the persistent
emission observed after type II bursts with durations between $\sim$
60 and 120 sec; in several cases the QPO evolved from $\sim$ 4 Hz just
after the burst to $\sim$ 2 Hz.  The frequency of the QPO was
positively correlated with the spectral hardness of the persistent
emission, such that the spectrum softened as the QPO frequency
decreased.  Occasionally, frequencies of $\sim$ 0.4 - 1.0 Hz were also
detected (\cite{stella88a}).  QPO of frequencies $\sim$ 0.04 Hz have
also been found to sometimes occur at the onset of the persistent
emission just after the post-burst dip, correlated with the occurrence
of a spectrally hard ``hump'' and with the appearance of 4 Hz QPO
(\cite{lubin92b}).

Dotani \etal (1990), using the X-ray detector LAC on {\it Ginga}
(\cite{turner89}), observed QPO from the type II bursts of the Rapid
Burster, and was able to pick out individual oscillations in the QPO,
and observed the slow change of the frequency of the QPO. 

All of the neutron-star low-mass X-ray binaries are a member of one of
two distinct groups called Z sources and atoll sources (\cite{hvdk89}
-- hereafter \hvdk).  Sources in both groups show a strong correlation
between their X-ray spectral and fast-variability properties.  In this
paper, we present a study of the X-ray timing and the spectral
behavior of type II bursts and persistent emission of the Rapid
Burster in light of the source classification of other LMXB.

A previous attempt to describe the fast-timing behavior of type II
bursts from the Rapid Burster in terms of the Z/atoll paradigm has
been made by Dotani \etal (1990).  They found that the timing behavior
(\ie\ QPO and noise) in the type II bursts are most similar to behavior
in the Normal Branch (NB) of Z sources (\hvdk); however, if the NB QPO
are explained by a near-Eddington process (see Van der Klis 1994 and
references therein), then the large range of burst peak fluxes in
which the 2-5 Hz QPO are observed may argue against a common origin
for RB burst QPO and NB oscillations.  

In Section 2, 
we describe the data used in the present
analyses.  In Section 3, 
we analyze the spectral and timing
behavior of the persistent emission. In Section 4, 
we
perform a similar analysis on the bursts.  In Section 5, 
we take a closer look at the 1985 observations to compare the bursts
and persistent emission behaviors with each other and with the observed
behaviors in other LMXB. In Section 6,
we discuss the
results of these analyses in the context of the Z/Atoll paradigm and
present our conclusions.

Throughout this paper, when we refer to ``burst numbers'' for bursts
observed during the August 1985 observations, we use the sequential
numbering of these bursts by Stella \etal (1988a) and Lubin
\etal (1992b).  The term ``bursts'' refers only to the type II bursts,
unless it is explicitly stated that we refer to type I bursts.   Also,
all power density spectra shown are normalized according to Leahy
\etal (1983). 

\Section{The Observations} 
\label{sec:obs}
The Rapid Burster was observed with the Medium Energy (ME) Argon
detector array (Turner, Smith \& Zimmerman 1981) of {\it EXOSAT} on
eight occasions.  We have analyzed six of these observations,
excluding those two (1983 August 7 and 1983 August 26) which were made
with 4U/MXB~1728-34 in the field-of-view (FOV).  Data were selected
which had high time resolution ($\approxlt 150$ ms) with at least 4
energy channels, and which were obtained when 1728-34 was completely
excluded from the FOV.  The selected data are indicated in
Table~\ref{tab:data}.  We chose to include the 1984 July 17
observation for completeness in the spectral part of this study,
despite the fact that Lubin \etal (1991) found no evidence for QPO in
the bursts in this observation.  Because there were no QPO in these
bursts, we did not analyze these data for the burst timing properties.

In the following, we briefly summarize some basic characteristics of
the RB during these individual {\it EXOSAT} observations, and give
references to previous work based on them.

\begin{itemize}
\item{\it 1983  August 15 }~ This observation has been  described by Barr \etal
(1987).  During this observation, the Rapid Burster was observed to
give off strong persistent emission and type I bursts, but no type II
bursts were observed.  For the first three and last five hours of this 24 hour
observation, the ME was pointed directly at the Rapid Burster, placing
4U/MXB~1728-34 at a collimator efficiency of 21\%. From 00:57 to 9:58,
the Rapid Burster was observed with a collimator efficiency of 28\%,
while 4U/MXB~1728-34 was excluded from the FOV.  During this period,
the Rapid Burster was observed with one-half of the ME array, while
the other half observed the (off-source) background.

\item {\it 1984 July 17}~  One half the detector array contained only
the Rapid Burster, the other one only 4U/MXB~1728-34, at transmission
efficiencies of $\sim$35\% and $\sim$40\%, respectively.  Thus, the
overall transmission efficiency (in reference to the full array) was
$\sim$17.5\% and $\sim$20\% for the Rapid Burster and for 1728-34,
respectively.  Background (off-source) data were taken from UT July
18, 06:27-07:34.  A total of 243 type II bursts were observed in this
observation. No persistent emission was detected, and the type II
bursting activity is characterized by mode II.  Results from this observation have been
presented by Lubin \etal (1991) and Tan \etal (1991).

\item {\it 1985 August 28-29}~ During the first 2.3 hours (with the
exception of a 15 min period; see Stella \etal 1988a) the transmission
efficiencies (referenced to the full detector array) for the Rapid
Burster and the nearby source 4U/MXB 1728-34 were $ \sim 41 $\% and $
\sim 100$\%, respectively.  After UT 17:45 1728-34 was excluded from
the FOV and the transmission efficiency for the Rapid
Burster was 48\%.  The persistent emission was highly variable.
Strong $ \sim $ 2 - 4 Hz and 0.04 Hz QPO (and occasionally $ \sim 0.4
$ - 1.0 Hz QPO) were observed in the persistent emission as well as $
\sim 2 $ - 5 Hz QPO in the type II bursts.  Background (off-source)
data were taken from UT August 28, 14:54 - 15:23.  During the entire
observation, one of the eight detectors was turned off.  Between UT
August 29, 1:00 and 9:35 (the end of the observation), one of the
seven operating detectors malfunctioned producing an artificial
flaring phenomenon.  This was identified by ``flaring'' types of
events in only one detector in the housekeeping data, which were not
seen in any of the other active detectors.  Some data during this
period could not be fully analyzed.  A total of 40 type II bursts were
observed during the observation.  Although the bursts were all very
long (90--680 sec), the burst activity is best described as mode I.
Results from this observation have been presented by Stella \etal
(1988a,b), Tan \etal (1991), and Lubin
\etal (1992b, 1993). 

\item {\it 1985 August 30-31} Between UT August 30, 18:28 and August 31,
00:10 and between UT August 31, 02:24 and 07:33, both half arrays were
pointing in the same direction, and the Rapid Burster was observed
with a transmission efficiency (referenced to the full array) of $
\sim 46 $\%; 4U/MXB~1728-34 was not in the FOV.  Background
(off-source) data were taken from UT August 30, 17:08 - 18:28 and from
UT August 31, 07:34 - 07:45.  In addition, one half array was pointed
at the background (off-source) between UT August 31, 00:10 and 01:10;
the other half was off-source between UT August 31, 01:10 and 02:15.
The frequencies of QPO observed during this observation are the same
as observed during 1985 August 28-29, except that the frequencies of
QPO in the bursts were in the range 2-3 Hz.  A total of 55 type II
bursts were observed during this observation. Although the bursts were
all very long (40--200 sec), the burst activity is best described as
mode I. (For details on these observations, see \cite{stella88a}, b;
\cite{tan91}; \cite{lubin92b}, 1993.) During the
entire observation, one of the eight detectors was turned off (the
same one as during the Aug 28-29 observations).  Results from this
observation have been presented by Stella \etal (1988a,b), Tan \etal
(1991), and Lubin \etal (1992b, 1993).

\item {\it 1985 September 10 and 13}~ The Rapid Burster, when
observed, was in the FOV of both half arrays with a transmission
efficiency (referenced to the full array ) of $\sim$43\%; 1728-34 was
outside the FOV.  371 type II bursts were observed on September 10,
1985 and 170 on September 13 1985.  QPO with centroid frequencies of
5-7 Hz were observed in bursts of duration 9-30 seconds. There was no
persistent emission detected in these observations. The type II
bursting activity was characteristic of mode II.  Background was taken
during the inter-burst periods. Results from these observations have
been presented by Tan \etal (1991) and Lubin \etal (1991).
\end{itemize}

\Section{The Persistent Emission}
\label{sec:pe}
Persistent emission (PE) was observed between type II bursts from the
Rapid Burster with {\it EXOSAT} during the observations on 1985 August
28-29, and 1985 August 30-31. On Aug 15, 1983 persistent emission was
detected, but no type II bursts were observed.  Persistent emission
was not detected during the Sept 10 \& 13 1985 observations, nor during
the July 17 1984 observations.

\SubSection{PE Spectral Analysis}
\label{sec:pespecan}
\SubSubSection{PE Spectral Analysis:  1983 August 15}
Figs.~\ref{fig:83pehidccd}a \& b shows Hardness-Intensity Diagrams
(HIDs) and Fig.~\ref{fig:83pehidccd}c shows the Color-Color Diagram
(CCD) of the persistent emission observed during the 1983 August 15
observation for a ``soft'' color (5.7~-~9.3 keV/2.0~-~5.7~keV) and a
``hard'' color (9.3~-~17.0~keV/5.7~-~9.3 keV).  The 1\sig~error bars
are indicated on each data point.  Each point represents an
integration time of 600 seconds.  Type I bursts were not included in
the HIDs, nor in the CCD.  All counting rates were corrected for
aspect and normalized to the full array area. As the count rates were
low, it was not necessary to correct them for dead time, which we
estimate is $\approx 1\%$.  While the persistent emission count rate
varies by $\sim$ 25\%, the hard color and the soft color vary $\approx$ 15\% ;
(Figs.~\ref{fig:83pehidccd}a \& b).  

The CCD (Fig.~\ref{fig:83pehidccd}c) does not indicate any
significant correlation between either the hard or soft colors with
the X-ray intensity; the hard and soft colors are themselves
correlated, however this is not surprising as they share an energy
band.  Using a Spearman rank-order correlation test, we find
probabilities of 34\%, 92\%, and 99.9\% for the data shown in
Fig.~\ref{fig:83pehidccd}a, b, and c, respectively.  The existence of
the correlation between the hard and soft colors indicates measurable
spectral evolution in the persistent emission of this data.  The {\it
general} trend of this evolution is a decreasing soft color and an
increasing hard color (referring to Fig.~\ref{fig:83pehidccd}c, the
motion in the CCD with increasing time is from the lower right corner
towards the upper left corner).  

Because the 15 August 1983 observation was taken with different energy
channel boundaries than those in the other observations, the energy
ranges are very different from those of other observations.  This,
unfortunately, complicates a comparison of CCDs and HIDs between this
and the other observations.

\SubSubSection{PE Spectral Analysis: 1985 August 28-29 \& 30-31}
During the course of the 30-31 August 1985 observation, the On Board
Computer (OBC) mode was changed, resulting in different selections of
energy channel ranges.  For a period while the OBC mode HER7
(hereafter, ``I7''; \cite{andrews85}) was active, the selected energy
ranges overlapped -- but did not exactly match -- the energy ranges
for the remainder of the observation.  We overcame this by correlating
contemporaneous data from earlier in the observation when both OBC
modes I7 and HER5 (hereafter, ``E5''; which had sufficient energy
resolution) were active.  Comparing the background-subtracted count
rates in the E5 energy ranges with the corresponding I7 energy ranges
(integrated over 20 sec), we were able to derive linear relations
between the E5 and I7 count rates.

The linear fits to the data in the three energy channels used in this
analysis were quite good, with little systematic deviation,
introducing an acceptably small systematic uncertainty ($\approxlt
7\%$).  In Table~\ref{tab:mb}, we show the parameters for linear
conversion of counts in the first three energy ranges of the form
$({\rm counts E5}) = m \times ({\rm counts I7}) + b. $

Figure~\ref{fig:85pehidccd} shows the HIDs and CCD of the persistent
emission as observed during the observations 1985 August 28-29 and
30-31 for a soft color (3~-~5~keV/1~-~3~keV) and a hard color
(5~-~10~keV/3~-~5~keV).  Each point represents an integration of 100
seconds.  A sample error bar is included in each panel.  All counting
rates (1-20 keV) have been normalized to the full array area and
corrected for background, aspect, and dead time (\cite{stella85}).
While the count rate changes substantially (up to a factor of 2; panels
a, b, d \& e), the hard and soft colors also vary, although not
systematically. Using a Spearman rank-order correlation test,
probabilities of spurious correlation of the magnitude observed in
these panels range between 95-98\%.  Fig.~\ref{fig:85pehidccd}c shows
that the hard and soft colors are uncorrelated, with a Spearman
rank-order correlation test probability of 92\%; however, the hard and
soft colors were correlated on Aug 30-31 (Fig.~\ref{fig:85pehidccd}f),
with a probability of being spuriously generated of $\approx 4\times
10^{-10}$.  This cannot be due to a systematic shift of the corrected
I7 data; we analyzed the E5 data separately from the corrected I7
data, and found that they separately have probabilities of $\approx
2 \times 10^{-4}$ and $\approx 1 \times 10^{-7}$ of being produced
spuriously.  Thus, there is a correlation in colors observed in the
Aug 30-31 data which is not observed in the Aug 28-29 data.

To illustrate the distribution of intensity and hardness in the
persistent emission, Figure~\ref{fig:85intdist} shows the distribution
of the average intensities (1 - 20 keV) and the hard and soft colors
of the persistent emission intervals (100 sec integrations) for both
observations. The spectrum of the persistent emission, observed during
the August 30-31 observation, is {\it harder} in the hard color and
{\it softer} in the soft color than that observed during August 28-29
(Figure~\ref{fig:85intdist}, bottom panel).  This indicates that the
3-5 keV range decreased in total fluence relative to the 5-10 keV and
1-3 keV range between these two observations, even while the
count-rates remained roughly constant.  The distributions of the hard
ratios on the different days are considerably distinct, whereas that
of the soft ratio is less so, and the intensity even less so.  This
indicates that, even while the countrates may remain constant, the
relative spectrum at high energy ($>$ 5 keV) varies with greater
magnitude than at low energies ($<$ 5 keV).

The weighted averages of the PE hard color (5 - 10 keV)/(3 - 5 keV)
and the soft color (3 - 5 keV)/(1 - 3 keV) of the August 28-29
observations are 0.765 and 2.052 and of the August 30-31 observation
are 0.824 and 2.017, respectively. The formal errors on these values
are $<$1\%.

\SubSection{PE Timing Analysis}
We have examined several timing features of the Rapid Burster
persistent emission using Power Density Spectra (PDS). These are (a)
the VLFN component, (b) $\sim$ 0.04 Hz QPO, and (c) QPO for centroid
frequencies in the range 2-6 Hz.  High Frequency Noise (HFN) is
discussed in the following section.  We parameterize these features by
fitting them, using a least-$\chi^2$ method, with the models:
\label{sec:fit}
\begin{eqnarray}
 P_{\rm Lor}(\nu, \nu_{\rm c}) & = & \frac{A_{\rm
Lor}}{\pi}~\frac{(\Gamma/2)^2}{(\nu -
\nu_{\rm c})^2 + (\Gamma/2)^2} \label{eq:lor} \\
 P_{\rm pl}(\nu)  & = & A_{\rm pl} ~\nu^{-\alpha} \label{eq:plaw} \\ 
 P_{\rm const} & = &  A_0 \label{eq:cons} \\
 P(\nu) & = &  P_{\rm const} + P_{\rm pl}(\nu) + \sum_{i=1}^{N} P_{\rm Lor}(\nu, \nu_i) \label{eq:totpower}
\end{eqnarray}

where $\nu$ is a frequency, $A_{\rm lor}$ is the area under a
Lorentzian peak integrated $ - \infty < \nu < \infty$, $\nu_c$ is the
centroid frequency, $\Gamma$ is the full-width at half-maximum (FWHM)
of the Lorentzian, $N$ is the number of fit QPO peaks (usually 1--3)
and $A_{\rm pl}$ and $A_0$ are constants.  $P_{\rm lor}$ is used for
modelling the QPO, $P_{\rm pl}$ is used for modelling the VLFN, and
$P_{\rm const}$ is used to model the Poisson level.

\SubSubSection{PE Timing Analysis: 1983 August 15} 
The variability of the persistent emission in the 1983 Aug 15
observation has been investigated by Barr \etal (1987).  For QPO with
a FWHM of \about 15\%, they found that ``At a 90\% confidence level
the upper limits to the RMS amplitude of the QPO ranged from \about
9\% to \about 12\% (depending on frequency)'' for frequencies in the
range 2-5 Hz.  A more general search for ``coherent and quasi-periodic
oscillations in the frequency range 0.06 to 64 Hz'' was performed, the
results of which were that ``The source power density spectrum did not
reveal any significant excess variability above Poisson noise.''

We have searched for excess Very Low Frequency Noise (VLFN) in this
data set.  We produced 12 FFTs of frequency range 0.00065-5.6 Hz,
using data in the 5.9-17 keV energy range, being careful to avoid
telemetry breaks and type I bursts.  As there was little spectral
evolution in the dataset ($<$15\%) we did not sort the FFTs by
hardness.  The FFTs were summed into a single PDS.  In
Fig.~\ref{fig:barrpds}, we show the PDS, rebinned logarithmically
(panel a) and linearly (panel b).  The evidence for VLFN (modelled as
a power-law plus a constant Poisson level) is marginal, with a 2\sig\ 
upper limit in the frequency range 0.00065-1.0 Hz of 2.0\%, and
requiring a very steep power-law slope ($\alpha$=3.3).  We also fit a
Lorentzian plus a constant to the PDS (rebinned linearly to 112 bins)
to search for QPO with FWHM \about 25\% at $\nu$=2, 3, 4, and 5 Hz.
We find no evidence for QPO; 2\sig~upper limits on the \%rms were in
the range 8 - 10\%, consistent with the upper-limits found on narrower
QPO by Barr \etal (1987).  Visual inspection of the PDS
(Fig.~\ref{fig:barrpds}b) shows no evidence for QPO-like behavior at
any frequency.
 
\SubSubSection{PE Timing Analysis: 1985 August 28-29 \& 30-31} 
\label{sec:85pdspe} 
To investigate the fast-timing properties of the persistent emission,
we performed a Fourier analysis using the high time resolution I5 and I7 data
in the energy range 1 - 20 keV (see Section~\ref{sec:obs}).

Successive fast Fourier transforms (FFTs) were made of each persistent
emission interval, beginning after the dip (see Introduction)
following each burst, when the 1-sec average count rate became 90\% of
the following PE mean count rate.  Each FFT was 128 sec long with a
Nyquist frequency of 16 Hz.  FFTs were made sequentially in each
interval until there was no longer enough data for an FFT before the
dip which precedes the next burst.  The resulting PDS (a total of 454)
were sorted according to their average hard color (5 - 10 keV/3 - 5
keV).  The PDS averaged into six hard color intervals are shown in
Figure~\ref{fig:pepds} with their least-$\chi^2$ models, from which
the strength of the spectral features (VLFN \& QPO) were found (see below,
and Tables~\ref{tab:peparams} \&\ref{tab:perms}).  The PDS have been logarithmically rebinned to
better show the VLFN component; the dead time affected Poisson level
(fit as a constant) has been subtracted.  The frequency bins where the
power was consistent with (less than 1\sig~greater than) zero are
plotted with 2\sig~upper limits (indicated by solid symbols and an
arrow pointing downwards); otherwise, the power in each bin is shown
with a 1$\sigma$ error bar.  The number of FFTs and the range of the
hard color of each averaged PDS are indicated in the upper right of
each panel.  The best fit model often lies below the plotted points,
as the model takes into account the data which are consistent with
zero power (here, plotted as 2\sig~upper limits).

The PDS of the persistent emission exhibit several distinct features
which vary with hard color.  As previously shown in Figure 9 of Stella
\etal (1988a), the frequency of the QPO increases from $\sim$ 2 Hz to
$\sim$ 4 Hz with increasing hard color
(Figures~\ref{fig:pepds}c,d,e,f).  For the 2-5 Hz QPO in the PE we
find, for the three softer color ranges, roughly constant \%rms values
of $\sim$ 8-11\%.  In the fourth hard color range (0.82-0.86), the QPO
\%rms increased (to 16.5\%) and the peak becomes much broader
(FWHM$=5.0\pm0.4$).  To account for multiple peaks in the PDS of the
two hardest ranges, we fit two Lorentzians to the data. To determine
the total \%rms under these double peaks, we summed the total power
under both Lorentzians.  We find that the \%rms is 17.1$\pm$1.9\% and
26.2$^{+1.7}_{-3.4}$\% in the two hardest color intervals.

Only the {\it harder} portions of persistent emission exhibit the
$\sim$ 40 mHz oscillations (Figures~\ref{fig:pepds}e,f), consistent
with the results found by \cite{lubin92b}.  The width of the 0.04 Hz
peak is systematically ill-constrained in panel f due to the poor
frequency resolution. No $\sim$ 0.04 Hz QPO are detected when the
persistent emission is softest.  A 2\sig~upper-limit to QPO in the
0.82-0.86 range is $\le 2.5\%$ (rms).  The \%rms increases in the
harder PE ($>$0.86 hard color) confirming the correlation with hard
color of 0.04 Hz QPO as first seen in this data by Lubin \etal
(1992b).  However, we are unable to significantly constrain the \%rms
(2\sig upper limit $\sim$84\%) in the hardest PE due to poor frequency
sampling.

The strength of the VLFN, integrating a power-law fit from 0.0078-1.0
Hz, shows a clear correlation in \%rms with hard color, increasing
from $<$2.9\% to 17.5\% with increasing hard color.  This may be due,
in part, to the appearance of several ``flare'' -type features in the
intensity profiles of this observation which have a very hard
spectrum; additionally, these may be due to the ``humps'', which
sometimes appear in the PE just after the dip following the burst, and
which are correlated with the appearance of the 0.04 Hz QPO and the
presence of a hard spectrum (see \cite{lubin92b}). It is important to
note that VLFN with a steep power law slope ($\alpha$ \approxgt 2.0)
is susceptible to low-frequency leakage (see \cite{vdk89}).

To investigate the contribution to the PDS by the ``hump'' (see
Introduction) in the PE, we excluded from the analysis the first 256
seconds of persistent emission following the start of the hump.  The
resulting PDS are shown in Fig~\ref{fig:nohumppds}, and were produced
as described above.  The least $\chi^2$ fit parameters are listed in
Table~\ref{tab:nohumpparams}, and the \%rms values of the different
features derived from these parameters are listed in
Table~\ref{tab:nohumpperms}.  The fits were largely poor, with very
high $\chi^2$ values.  

It is clear that the low frequency power is diminished when the hump
is excluded, and is therefore probably due to the hump itself (and
thus, is not VLFN).  The power-law slopes were, in some cases,
unmeasurable.  The strength of the low frequency power increased with
increasing hardness (from $\le$ 2.2 to 7.5$^{+1.4}_{1.8}$\%rms),
however dropping in the very hardest range to 3.5$\pm0.6$. 

The exclusion of the ``hump'' has little qualitative or quantitative
effect on the dependence of QPO parameters on spectral hardness.

The strength of the QPO with the first 256 sec following bursts
excluded (Table~\ref{tab:nohumpperms}) is nearly identical as a
function of spectral hardness to strength of the QPO when the first
256 sec are included (Table~\ref{tab:perms}).  The QPO \% rms
(0.01-100 Hz) as a function of spectral hardness is largely constant,
with a small dip in the middle spectral range and a jump at the
hardest spectral range.

{\it High Frequency Noise.} Using the available HTR3 (hereafter,
``T3'') data, we have searched for High Frequency Noise (HFN) above 10
Hz in the persistent emission.  The T3 data has very high time
resolution ($\le$ 4 ms) with no energy resolution (1 channel; $\sim$ 1
- 22 keV).

Data were available for 36 persistent emission intervals (i.e.  36
between-burst intervals) during the August 1985 observations.  Each
interval was divided up into consecutive 128 second stretches, each of
which produced a single FFT.  The FFTs were averaged together into a
single PDS for each persistent emission interval, and the Poisson
level, calculated as described in Van der Klis (1989), was
subtracted. The \%rms value was determined in two frequency ranges:
10-100 Hz and 20-60 Hz.  These were selected because they are the frequency
ranges where the HFN is dominant in the Z-sources and where the QPO in
the horizontal branch occurs in the Z-sources (\hvdk).  The \%rms was
measured by summing the Poisson-level subtracted power in the relevant
frequency range.  Of the 36 PE intervals, one (following burst
\# 41) had measured HFN in the 10-100 Hz frequency range
at a significant level ($18\pm3$\%).  One PE interval (following burst
\# 58) had a significant  \%rms (16 $\pm$ 3\%) value in the frequency range 20-60 Hz. The
other 35 PE intervals had 2\sig~upper limits to their HFN in the range
of 15\%-62\%, with most upper-limits below 30\%.  With all FFTs from all 36
PE intervals averaged into one PDS, the 2\sig~upper limits to the HFN
in the 10-100 Hz range and 20-60 Hz range were 8.0\% and 10.0\%,
respectively.

We also sorted these persistent emission FFTs by their hardness and
produced averaged PDS.  No hard color range produced a detection of
HFN, in either the 10-100 Hz or 20-60 Hz ranges, with 2\sig~upper
limits of 13-22\% and 10-17\%, respectively.

\Section{The Type II Bursts}
\label{sec:bursts}
\SubSection{Spectral Analysis}
\label{sec:burstspec}
Figure~\ref{fig:hidccdcomp} shows HIDs and CCDs of all five
observation periods in which type II bursts were observed from the
Rapid Burster (1984 July 17, 1985 August 28-29, 1985 August 30-31,
1985 September 10 \& 13).  The observation period is indicated at the
top of each column.  Each data point represents all data from a single
burst.  For 1984 July 17 and 1985 September 10 \& 13 observations,
when the bursts were relatively short (``short-burst observations'')
data were integrated from the first point which is $>$5\sig~above
background to the first point which is 0.5\sig~below background.  For
the ``long-burst observations'' (1985 August 28-29 \& 30-31), data were
integrated from the first of two consecutive time bins in which the
count rate is greater than 6$\sigma$ above the pre-burst persistent
emission level, and the burst end is defined as the first of two
consecutive time bins in which the count rate is within 2$\sigma$ of
the post-burst persistent emission level (excluding the dip).  The
difference between the burst-start and burst-end definitions is due to
the fact that long bursts have many more integrated counts than the
short bursts and we can afford greater precision in delimiting their start
and end; short bursts, however, must be delimited with the requirement
of maximizing the integrated signal/noise.

Each panel in Fig~\ref{fig:hidccdcomp} contains an average 1$\sigma$
error-bar, located at the same coordinate along the same row. These
error bars are not located at actual data points.

Figure~\ref{fig:hidccdcomp} shows the burst peak count rate vs.  burst
hard color (5-10 keV/3-5 keV; panels a-e) and vs. burst soft color
(3-5 keV/1-3 keV; panels f-j).  The peak count rate is the highest
$\sim$ 1 sec time-bin count rate during the burst, and has been
corrected for background (see Section~\ref{sec:obs}), dead time (see
\cite{stella85}; eq. 3.10 of \cite{vdk89}), aspect, and renormalized
to the full detector array area.  Background was subtracted from the
channel count rates prior to calculating the colors. Peak flux was
selected as a discriminator as the spectral properties of the bursts
are a function of the burst peak flux (\ie\ brighter bursts are harder,
and have lower QPO centroid frequencies than fainter bursts; see
Introduction).

Figs.~\ref{fig:hidccdcomp}k-o show the CCDs of soft color
(3-5~keV/1-3~keV) vs. hard color (5-10~keV/3-5~keV).  Immediately,
one's eye is caught by the extended branches observed during the
long-burst observations.  This is most evident in the CCD
(Figs.~\ref{fig:hidccdcomp}l,m), but it is also evident in the soft
color HIDs (Figs.~\ref{fig:hidccdcomp}g,h) and the hard color HID
(Fig.~\ref{fig:hidccdcomp}b; the hard-color HID in
Fig.~\ref{fig:hidccdcomp}c shows more an amorphous and extended
``cloud'' than a branch).

These well-defined branches stand in contrast to the ``clouds''
observed during the other three observation periods (1984 July 17,
1985 Sept 10, and 1985 Sept 13;
Figures~\ref{fig:hidccdcomp}a,d,e,f,i,j,k,n,o).  The peak count rates
and the dynamic range of the peak count rates during the long-burst
observations are larger (by a factor of $\sim$1.5 in peak count rate,
and $\sim$ 1.5-2.2 in dynamic range) than those observed during the
short-burst observations.

The previously reported correlation between burst peak counting rate
and spectral hardness (or equivalently blackbody temperature) for long
($> 30$ sec) type II bursts (\cite{kunieda84}; \cite{tan91};
\cite{lubin92a}) is clearly evident in the HIDs of both observations.

The bursts observed during the August 28-29 observation have a larger
dynamical range in hardness than those observed on August 30-31.
Bursts with relatively low average peak intensities ($\sim$ 1200 -
1400 counts/s; 1 count/s $\simeq$ 1.6 $\times$ $10^{-11}$ ergs
cm$^{-2}$ s$^{-1}$; see Tan \etal 1991) and therefore with softer
spectra were only observed during the August 28-29 observation. The
weighted averages of the hard color (5 - 10 keV)/(3 - 5 keV) and the
soft color (3 - 5 keV)/(1 - 3 keV) of the August 28-29 observations
are 0.94 $\pm$ 0.01 and 2.93 $\pm$ 0.03 and of the August 30-31
observation are 0.99 $\pm$ 0.01 and 2.98 $\pm$ 0.01, respectively (see
Table~\ref{tab:avghardratios}).

Table~\ref{tab:avghardratios} gives the weighted mean burst hard (5-10
keV/3-5 keV) and soft (3-5 keV/1-3 keV) colors for these five
observations. The spectral state of the 1984 July 17 cannot be
directly compared to that of the 1985 observations, as changes in the
EXOSAT ME calibration can account to up to a 10\% shift in the
hardness ratios in these channels (see below).  Comparing the 1985
observations, however, the hard ratio is lowest  during the 1985 Sept
13 observation (0.892 $\pm$ 0.003).

By averaging the short-burst colors and intensities some structure is
revealed in the ``clouds'' of Fig.~\ref{fig:hidccdcomp}.  The
composite HIDs and CCDs of Fig.~\ref{fig:burstcomp} illustrate this.
The observation period is indicated by the legend in the figure.  For
the short-burst observation periods, we plot the average hard and soft
colors of bursts with 1-20 keV peak count rates within consecutive peak
count rate ranges; these ranges are indicated as error bars in peak
count rate.  The error bars on hardness for July 1984 and September
1985 are the 1\sig~uncertainties in the mean hardness.  Each point
from the long-burst observations represents data from a single burst
($\approxgt$ 40 s).  The average 1\sig~uncertainty is indicated as an
error bar on a single point for each of these two observation periods.
All count rates have been corrected for dead time, aspect, and
normalized to the full EXOSAT ME array area.

In both the HIDs and the CCD in Figure~\ref{fig:burstcomp}, the
July\label{sec:speccomp} 1984 bursts are separated from the other
bursts, occupying a completely different area on the HIDs and CCD.  We
investigated if this separation could be due to the change in the
EXOSAT ME energy response.  We assumed a detector energy response
which was Gaussian, with a centroid energy, detector efficiency, and
FWHM as given by the relevant EXOSAT calibration file.  We assumed a
source thermal spectrum with a galactic absorption component with
$N_h = 1.5 \times 10^{22}$ cm$^{-2}$ (see, for instance,
\cite{lubin92a}).  We did this for three separate temperatures, kT=
1.5, 1.85, 2.2 keV.  We find a systematic shift in the soft ratio of
$<$1\% and a systematic shift in the hard ratio of \approx 10\%
between the 1984 and 1985 calibrations.  The shift in the hard ratio
due to the calibration shift of EXOSAT can completely account for the
separation in the hard ratio between 1984 and 1985 observations seen
in Fig~\ref{fig:burstcomp}b.  However, it cannot account for the shift
in the soft ratio in Fig.~\ref{fig:burstcomp}c.

\SubSection{Timing Analysis}
\label{sec:timing}
In this section, we present the results of a timing analysis of bursts
in the August 1985 and September 1985 observations.  We did not
analyze the July 1984 observation for timing properties, as a previous
study (\cite{lubin91}) observed no QPO in these bursts. To attempt to
deconvolve the LFN from the burst profile is not useful, given that
the burst durations were short (largely 3-12 seconds) and most of the
broad-band structure is due to the burst profile (see \cite{tan91}).

\SubSubSection{Burst Timing Analysis: 1985 August 28-29 \& 30-31}
\label{sec:85pdsbursts}
To investigate the fast-timing properties of the bursts we performed a
Fourier analysis using the high time-resolution I5 (HER5 data-type,
but with no energy resolution), I7, and T3 data in the energy range 1
- 20 keV.

{\it Power Density Spectra.} Successive FFTs were made of each burst
peak between the peak start and end. To avoid the sharp rise of the
burst, the peak start was somewhat arbitrarily defined as the first in
three consecutive 1-sec time bins where the average count rates in the
following two bins are no more than 2$\sigma$ greater than that in the
first bin.  The peak end was defined as the first bin in the first
series of two bins (1 sec integration each) in which the count rates
were both more than 2$\sigma$ below the average counting rate up to
that point; the characteristic ringing and humps (\cite{tan91}) are
thus not analyzed.  The shortest burst peak (defined in this way) was
18 seconds.  Each FFT was 16 sec long (with a Nyquist frequency of 16
Hz); therefore, each burst consisted of one or more FFTs depending on
the burst duration.  The resulting FFTs (a total of 455 for both
observations) were sorted according to the hard color (5 - 10 keV/3 -
5 keV) of the PDS interval (not according to the hard color of the
entire burst).  Figure~\ref{fig:burstpds} shows the average PDS of
these six hard color intervals.  The number of FFTs and the hard color
range of each averaged PDS are indicated in the upper left of each
panel.  The dead time affected Poisson level has not been subtracted.
The solid line in each panel is the least-$\chi^2$ best fit model,
found as described below (see Tables~\ref{tab:burstparams}
\& \ref{tab:burstrms}).

{\it QPO Frequencies.} The dominant feature in several of the averaged
power density spectra is a QPO peak.  The QPO
frequency decreases from $\sim$ 5 Hz for the softest bursts to $\sim$
2 Hz for the hardest bursts.  For PDS of the softer (0.74-0.98) periods,
the mean QPO frequency is measured to be 4.4 - 4.9 Hz, while for the
harder periods (0.98-1.12), the QPO frequency is measured to be 2.0 -
2.5 Hz (see Table~\ref{tab:burstparams}).  QPO at frequencies 2-5 Hz
manifest themselves in the hardness range 0.92-0.98 as a broad peak
($\Gamma$ = 1.7 Hz).  Though the average PDS of the bursts seem to
show QPO frequencies of $\sim$ 2 and $\sim$ 5 Hz
(Figure~\ref{fig:burstpds}), the centroid frequencies in individual
bursts actually extend almost uniformly from $\sim$ 3.5 - 5 Hz and
from $\sim$ 2 - 2.5 Hz (see Fig.  3 of Stella \etal 1988a).  The {\it
general} anti-correlation between QPO frequency and burst hardness is
expected as the frequency of QPO in long ($> 30$ sec) bursts is
anti-correlated with the burst peak intensity (\cite{stella88a}), and
the burst peak intensity is correlated with hardness
(\cite{kunieda84}; \cite{tan91};
\cite{lubin92a}; Figure~\ref{fig:HID85}).

{\it Fractional rms Variations.} Table~\ref{tab:burstrms} shows the
\%rms variation (dead time and background corrected) in the VLFN
component and the QPO($\nu$=2-5 Hz) as a function of the hard color (5
- 10 keV)/(3 - 5 keV).  These values were derived from the average PDS shown
in Figure~\ref{fig:burstpds}.  While the QPO frequency decreases with
increasing hard color, the \%rms first decreases, then increases (see
Table~\ref{tab:burstrms}). There is a very strong anti-correlation
between the hardness and the fractional rms variation of the QPO for
the softest intervals,
which inverts to a {\it correlation} for the  hardest ranges, where
the QPO frequency is $\sim$ 2 Hz.  Fractional rms variations extend 
up to $\sim$ 26\%.

The VLFN component is represented by the best-fit power law (see
Table~\ref{tab:burstparams} and Eqs.~\ref{eq:lor}-\ref{eq:totpower}).  The
\%rms values (integrated between 0.06-1 Hz) for all 6 hard color ranges
are consistent with a constant value ($\sim$ 2\%), except for the
softest two intervals, where the 2\sig~upper-limits are $\le$ 2.5\%~
and $\le$ 2.4\% (Table~\ref{tab:burstrms}), which were found by fixing
the power law slope to 2.00. 

{\it High Frequency Noise.} Data of type T3 were available for 34
bursts during the August 1985 observations.  Each of these 34 burst
peaks was divided in 16 second intervals, as described above.  The
FFTs were averaged together into a single PDS for each burst, and the
Poisson level (fit as a constant) was subtracted.  The \%rms values
were measured by summing the Poisson-level subtracted power for two
frequency ranges: 10-100 Hz and 20-60 Hz.  No significant HFN was seen
in either the 10-100 Hz or 20-60 Hz frequency ranges.  For the 10-100
Hz range 2\sig~upper limits were 6-12\% with most values below 9\%;
for the 20-60 Hz range, $2\sigma$ upper limits were 5-10\% with most
values below 8\%.  When all FFTs from all bursts are summed into a
single PDS, we find a \%rms $2\sigma$ upper limits of 3.0\% and 3.6\%
in the 10-100 Hz and 20-60 Hz ranges, respectively.

We also separated the burst FFTs based on their instantaneous hardness
and produced averaged PDS in the same hardness intervals used above.
The FFTs in each hardness range were averaged together into a single
PDS, and the Poisson level, calculated as described by Van der Klis
(1988) was subtracted.  The \%rms values were calculated by summing
the Poisson subtracted power over the relevant frequency range.  The
softest burst interval (hardness 0.74-0.80) had HFN with \%rms
$9.4\pm2.7$ and $10.8\pm3.6\%$ in the 20-60 Hz and 10-100 Hz frequency
ranges, respectively, which we interpret as only marginally
significant.  No other individual energy spectral range produced a
significant \%rms, with typical 2\sig~upper limits in the range of
4-6\% and 4-7\% for the 20-60 Hz and 10-100 Hz frequency ranges,
respectively.

\SubSubSection{Burst Timing Analysis: 1985 September 10 \& 13} 
{\it Power Density Spectra.} We have investigated the timing behavior
as a function of peak count-rate of the bursts observed during the 1985
September 10 \& 13 data.  Of the 541 bursts in this data set, 312 were
observed using the high time resolution (2-4 ms) T3 OBC mode during
periods without telemetry loss.  These bursts were separated into
three duration groups: $\tdur < 8{\rm s},~8{\rm s} <
\tdur < 16{\rm s},$ and $ 16{\rm s} < \tdur < 32{\rm s}$.  There were no bursts longer
than 32 seconds.  For the purpose of this timing analysis, the burst
``start'' was defined as the first time bin which was $5\sigma$ (or
greater) above the pre-burst background count rate; the burst ``end''
was defined as the first time bin that was $0.5
\sigma$ {\it below} the pre-burst background count rate.

This criterion is different from that used to determine the period
over which FFTs were taken for the August 1985 observations because of
the very different time-evolution of the bursts during these
observations; while the bursts observed during August 1985 were long
($40-680{\rm s}$) and relatively flat-topped, the bursts during the
September 1985 observations were short ($<32 {\rm s}$) with a slowly
decaying and ``ringing'' intensity profile (see \cite{tan91}).  Thus,
we include the burst decay in the timing analysis of the September
1985 observations, while we could afford to exclude the burst decay in
the timing analysis of the August 1985 observations.  We caution that
the shape of the PDS below $\sim$ 1 Hz can be a strong function of the
choices of ``burst start'' and ``burst end'' for a burst with a slowly
decaying intensity profile, because the overall burst profile produces
significant power at low frequencies.

After the bursts were separated according to their duration, one FFT
was taken of each burst.  The start-time of each FFT was taken 1
second after the burst ``begin'' (see above) to ensure that the rise
of the burst, which can cause spurious spikes in the PDS, was not
included in the FFT.  Bursts of $\tdur < 8{\rm s}$ had 8s FFTs; bursts of duration $8{\rm s}<
\tdur < 16{\rm s},$ had 16s FFTs; and bursts of duration $16{\rm s} <
\tdur < 32{\rm s}$ had 32s FFTs.  The Nyquist frequency for all
FFTs was 128 Hz.  Because the QPO frequency varies strongly with peak
count rate (\cite{lubin91}), the FFTs were then grouped by the peak
count rate (see Figures~\ref{fig:hidccdcomp}d, e, i, \&j) into three
groups with roughly equal numbers of bursts in each: $\ip < 1000{\rm
c/s},~1000 < \ip < 1100 {\rm c/s}$, and $ 1100 < \ip < 1500 {\rm
c/s}$.  PDS in the same countrate range but with different time
durations were rebinned to the coarsest frequency resolution.  These
rebinned PDS were then averaged into a single PDS for each peak count
rate range. 

We account for differences between power due to random processes (\eg\ 
noise features, QPO) and power due to deterministic processes (\ie\ 
the burst envelope) by finding the frequency below which deterministic
processes dominate the power spectrum and above which random process
dominate. Fig.~\ref{fig:notcesme} shows results of an average PDS of
91 FFTs from bursts with duration $8{\rm s} < \tdur < 16{\rm s}$ and
peak count rate $ 1000 < \ip < 1100{\rm c/s}$.  Panel {\it a} shows
the average power as a function of frequency.  Panel {\it b} shows the
ratio of $\sigma / P $ as a function of frequency; here, $\sigma^2 =
(<P^2> - <P>^2)/{\rm NFFTs}$ is the squared uncertainty in the mean
power of all 91 FFTs as a function of frequency, and $P$ is the mean
power as a function of frequency.  Panel {\it c} shows $\sigma$ as a
function of frequency.  Down to frequencies of about 0.5 Hz, $\sigma
\times \sqrt{\rm NFFTs}$ is equal to the average power, indicating
that the power spectrum is dominated by a random process.  At lower
frequencies, $\sigma$ falls below this, as here the timing properties
of the bursts are dominated by the deterministic burst envelope
(see~\cite{vdk89}).

We produced averaged PDS of the FFTs grouped by peak count rate, and
one averaged PDS of all 312 FFTs (Fig.~\ref{fig:985totpds}).  In our
fits to these PDS, we assumed a constant dead time induced Poisson
level, which has been subtracted from the data, and rebinned the
resulting PDS logarithmically.

{\it Fractional rms Variations.}  Because the PDS below $\sim$ 0.6 Hz
are affected by the burst envelope, we used a two step modelling
procedure (see Section~\ref{sec:fit},
Equations~\ref{eq:lor}-\ref{eq:cons}) to separate the power due to the
deterministic burst envelope and the random process.
\label{sec:985pdsmodel} In the ``primary'' fit, we used the model
$P(\nu) = P_{pl}(\nu) + P_{\rm const}$ to fit the data at frequencies
$\nu <0.6$ Hz where the power is dominated by deterministic power and
$\nu > 20$ Hz, where the power is assumed to be Poisson.  We had
attempted to subtract the deadtime-affected Poisson level, using the
method described by Van der Klis (1989).  Because of the significant
intensity decay of the burst, the ``constant count rate'' assumption
in the method described by Van der Klis (1989) is no longer valid, and
we found that the Poisson level was over-estimated. We thus decided to
assume a constant Poisson level and no measurable HFN.  In the second
step, we performed a ``residual'' fit, with the model $P(\nu) = P_{\rm pl,
  fixed}(\nu) + P_{\rm const, fixed} + P_{\rm pl}(\nu) + P_{\rm
  Lor}(\nu, \nu_{\rm c})$ to fit the data at frequencies $ 0.6 {\rm
  Hz} < \nu < 20$Hz, where $P_{\rm pl, fixed}(\nu)$ and $ P_{\rm const,
  fixed}$ are the fixed values of the model from the ``primary'' fit.
Figure~\ref{fig:985totpds} shows the 4 averaged PDS with their
``composite'' fit (\ie~the sum of both primary and residual fit).

The parameters derived from the primary and residual fit are given in
Table~\ref{tab:sept85burstfitparams}.  The $\chi^2$ values are very
high; the fits are poor, due largely to poor fitting in the 0.7-3 Hz
frequency range.  We attempted to fit models to the data with the
0.7-3 Hz range excluded, but the fits became highly unstable due to
the interation between the power-law and the Lorentzian parts of the
fits.  The primary power-law slopes are constant, independent of the
peak count-rate, at $\alpha \approx 3.1$.  The ``residual'' power-law
slope in the lowest peak count-rate range was unconstrained, and
destablized the fit, so we fixed it to a mean value of the other
residual power-law slopes (which were in the range $0.7$-$0.9$) to
obtain an upper-limit on the ``residual'' VLFN power.

We confirm the previously reported correlation of the QPO centroid
frequency with peak count rate in this peak intensity range found by
Lubin \etal~(1991) for this data set; however, note that at higher
peak intensities, the relation has been found to be an {\it
  anti-}correlation (Tawara \etal 1982; Stella \etal 1988a; Lubin
\etal 1991).  Also, in these bursts, there appears to be an
anti-correlation between QPO centriod frequency (or, burst peak
count-rate) and QPO FWHM, which we measure as decreasing between the
three count-rate ranges from 5, to 2, to 1 Hz.  This anti-correlation
is not globally observed from bursts in the Aug 28-29 and 30-31 1985
observations, with QPO frequencies in the range of 2-5 Hz (above) and
Stella \etal (1988a) find a correlation between QPO centriod frequency
and QPO FHWM in the individual Aug 1985 bursts.

The \%rms values derived from the models are presented in
Table~\ref{tab:sept85burstrms} and have been corrected approximately
for background, deadtime, and burst-duration, and changing burst
intensity.  We see no correlation between the burst peak intensity and
the \%rms values in either the primary power-law model, the residual
power-law model.  The sum of the primary and residual \%rms is
constant with increasing peak count rate at $\approx$ 50\%.  The \%rms
of the QPO appears to be anti-correlated to peak intensity decreasing
from 21\%, to 15\% to 7\% in the three different intensity ranges.
However, we strongly caution against interpreting this data as
indicating the existence of such an anti-correlation, due to the large
systematic uncertainty caused by the extremely poor fits of the model
to the data ($\chi^2/\nu \approx 10$).  We attempted to overcome this
by excluding the frequency region responsible for the poorness of
these fits (0.6-2 Hz, due to the burst ``ringing''), however, this
left the residual VLFN largely unconstrained.  If we exclude the VLFN
from these fits, the QPO \%rms became very high, as the QPO FWHM
broadened to compensate for the absence of a model at the
low-frequency range.  This systematic uncertainty accounts for the
discrepancy between this result and Lubin \etal (1991), where the
\%rms of the QPO for bursts, separated by duration, were \approxlt
10\% for about 2/3 of the duration ranges, and in the range 12-24\%
for the remaining 1/3 of the duration ranges.

\Section{The August 1985 Observations}
\label{sec:aug85}
Because of the richness of the 1985 August 28-29 \& 30-31 observations
we summarize and compare the spectral results of the bursts and PE in
these data sets. 
\SubSection{Spectral Results}
\label{sec:aug85spec}
\SubSubSection{HIDs} 
\label{sec:aug85hid} 
Figure~\ref{fig:HID85} shows hardness-intensity diagrams (HIDs) of the
bursts and the persistent emission for a soft color (3 - 5 keV/1 - 3
keV) (Figures~\ref{fig:HID85}a) and a hard color (5 - 10 keV/3 - 5
keV) (Figures~\ref{fig:HID85}b).  The data points
from the August 28-29 and August 30-31 observation are indicated with
filled and open squares, respectively.  For the burst data, each point
in the HIDs represents an integration from the burst start to the
burst end (see Section~\ref{sec:burstspec}). The bursts vary in
duration from $\sim$ 40 to 680 sec.  For the persistent emission, each
point represents an integration of 100 sec.  The intensities are the
{\it average} counting rates (1 - 20 keV) of each integration
interval.  All count rates have been normalized to the full array.
Corrections for collimator transmission (aspect), background, and dead
time (see \cite{stella85}) have also been made.

In the HIDs, the bursts lie considerably away from the persistent
emission, largely because of the difference in source intensity.
There is also a great spectral difference between the bursts and the
persistent emission, which is most plain in the soft ratio (3-5
keV/1-3 keV; Fig.~\ref{fig:HID85}a), where the bursts lie above the
range of values occupied by the persistent emission.  The spectral
difference is not as great when considering the hard ratio
(Fig.~\ref{fig:HID85}b), where there is considerable overlap in the
parameter space of the bursts and persistent emission. 

\SubSubSection{CCDs}
\label{sec:aug85ccd}

Color-color diagrams (CCDs) were made of both the bursts and the
persistent emission for each observation (Figure~\ref{fig:85CCDcomp});
all data from the two observations were used except when the nearby
source 1728-34 was in the FOV (see Section~\ref{sec:obs}).  The time
intervals used to calculate the colors of the bursts and the
persistent emission were the same as those described in Sec. 4.1.

In the CCD, the bursts clearly lie away from the PE, due largely, but
not exclusively to the great difference in soft-ratio (3-5 keV/5-10
keV).  The locus of of CCD points of the bursts is an elongated track,
with one end roughly at the median hard-ratio of the persistent
emission, but with a measureably higher soft-ratio, demonstrating the
harder spectrum of the bursts compared to the persistent emission.

As with the persistent emission, the average burst hard color on
August 30-31 is higher than on August 28-29.  However, when one
considers the soft color, the trends are {\it opposite} (\ie\ the PE
soft color is lower, while the burst soft color is
higher).\label{sec:colorbursts}

\SubSection{Relation between Burst and Persistent Emission Hardness}

We have investigated the possibility that the hardness of the bursts
is correlated to that of the following persistent emission.
Figure~\ref{fig:burstpe} shows the soft color (3 - 5 keV/1 - 3 keV)
and the hard color (5 - 10 keV/3 - 5 keV) versus burst number. A solid
point indicates the hardness of a burst while an open point indicates
the hardness of the {\it following} persistent emission interval.
(Some burst and persistent emission data could not be included due to
telemetry losses or contamination by 1728-34.)  A {\it global}
correlation between burst and persistent emission hardness is most
evident in the hard color (Figure~\ref{fig:burstpe}b) where there is a
clear drop in the hardness of both the bursts and the persistent
emission between bursts \# 28 and 40.  (These bursts have the lowest
peak counting rates and therefore the softest spectra; see
Figures~\ref{fig:HID85}a,b.)  The drop can not be attributed to the
detector malfunction which occurred at burst \#28, as the relevant
experiment half was excluded from analysis; we also compared hardness
ratios throughout the entire observation produced with only the
experiment half which did not malfunction, and the global relation is
still present.

There also exists a correlation between the burst spectrum and the
persistent emission immediately following. 

In Fig.~\ref{fig:hardbpe}, we show the burst hard color vs.  that of
the following PE, and the burst soft color vs.  that of the following
PE, separately for the Aug 28-29 (bursts 1-39) and Aug 30-31 (bursts
40-95) observations.  Using a Spearman rank-order correlation, we find
that, there exists a linear correlation between the burst hard and
soft ratios that of the following persistent emission burst hard and
soft ratios, with probabilities of such correlations being spuriously
generated of $10^{-4}$ -- $10^{-3}$.  Thus, on both days, there is a
measured correlation between the soft- and hard- color of the bursts
and the following persistent emission on both days. Specifically, the
correlation of the soft ratio of the burst and the following
persistent emission is significant at 99.87\% on Aug 28-29, and at
99.94\% on Aug 30-31.  The correlation between the hard ratio of he
burst and the following persistent emission is significant at 99.96\%
on Aug 28-29, and at 99.99\% on Aug 30-31.  This demonstrates that the
spectral hardness of the bursts is correlated with the following
persistent emission.

We have shown (Sec.~\ref{sec:colorbursts}; see Fig.~\ref{fig:HID85} \&
Fig.~\ref{fig:85CCDcomp}) that, averaged over the entire observation,
the PE and burst hard ratios both increase between Aug 28-29 and
30-31, but that the soft ratios follow opposite trends (decrease in
the PE, increase in bursts).  This may indicate that the emission of
the PE and bursts in the 3-10 keV range are globally correlated, while
that in the 1-3 keV range are either not correlated or possibly
anti-correlated.

\SubSection{Comparison of the Strength of Variability in the
Persistent Emission and Bursts} 
\label{sec:rmscomp}
To graphically illustrate the differences between the timing
properties of the PE and the bursts observed during the 1985 August
observations, we plot in Fig.~\ref{fig:bupermscomp} the \%rms values
of the 2-5 Hz QPO in the bursts (see Section~\ref{sec:85pdsbursts};
Table~\ref{tab:burstrms}) and the PE (see Section~\ref{sec:85pdspe}l;
Table~\ref{tab:perms}) as a function of hardness ratio.  The \%rms of
the QPO in the PE is generally correlated, while that of the bursts is
anti-correlated.  As for the VLFN, the \%rms first increases, then
decreases with decreasing hardness in the PE (see
Table~\ref{tab:perms}); in the bursts, however, the VLFN is constant
in the harder ranges, and below detection in the softer ranges (see
Table~\ref{tab:burstrms}).

\SubSection{Relation between QPO Frequency and Burst Position on the
HID} 
In Figure~\ref{fig:z}, using the HID in Figure~\ref{fig:HID85}b, we
marked the location of each burst with an approximate QPO peak
frequency, which was identified in averaged PDS of individual bursts.
The frequency was estimated ``by eye'', which required that a peak
appeared to be significantly above the Poisson level in the 2-7 Hz
range. This figure should be taken only to indicate somewhat
qualitatively the timing behavior of bursts as a function of position
on the HID.  Bursts observed on August 28-29 are indicated by a
circle, and bursts observed on Aug 30-31 are indicated by a square.
Bursts where no QPO peak was identified are indicated on the HID with
a dot inside the circle or square, and the circle or square was left
empty when no PDS was available (\ie\ the time resolution was poor or
the burst was interrupted).

Motivated by what appears like two separate CCD branches, we somewhat
arbitrarily separated the bursts on the HID into 5 separate groups
indicated by the letters A-E.  We averaged the FFTs from the bursts in
these groups (produced as described in Section~\ref{sec:85pdsbursts})
into single PDS, shown in Figure~\ref{fig:zpds}.  Groups B \& C, which
correspond to roughly the same hardness range, have roughly identical
PDS, with a small amount of VLFN and a sharp QPO peak at $\sim$ 2 Hz.
Groups A \& D, which also correspond to roughly the same hardness
range, have roughly identical PDS, with a small amount of VLFN and
little or no indication of QPO in the 2-7 Hz range.  Group E, the
softest group, has a different QPO peak frequency ($\sim$ 5
Hz) than seen in the other groups.  Thus, the PDS appear to be
determined by the hardness of the burst, rather than the position on
what appear to be branches.

We investigated the possibility of HFN correlated with position in the
CCD.  Using the available T3 data, we produced averaged PDS of FFTs
from bursts in groups A, B, C, D, E, A+B, and D+E (see above).  We
attempted model fits to the PDS above 10 Hz using $P(\nu) = A_{\rm
HFN}~\nu^{-\alpha}
\exp({-\nu/\nu_{\rm cut-off}})$ (see \hvdk) but found the fits were
unstable.  We instead averaged the power in the ranges 20 Hz $< \nu <$
60 Hz and 10 Hz $< \nu <$ 100 Hz for all 7 PDS.  No HFN was detected
with 2\sig~upper limits in the range 3-6\%. The limits on the HFN in
the PDS of A+B and D+E were 5.0\% and 3.4\% in the 10-100 Hz range and
4.7\% and 4.1\% in the 20-60 Hz range.

\Section{Discussion and Conclusions}
\label{sec:con}

Based on the combined X-ray spectral and fast-variability
characteristics, two types of NS LMXB can be distinguished,
the atoll and Z sources (\hvdk). 

Atoll sources exhibit two states.  The banana state is characterized
by an upwardly curved branch in the CCD and power density spectra
dominated by VLFN.  The island state is characterized by lower X-ray
intensities and little correlated motion in the CCD; its power density
spectra are dominated by high frequency noise (HFN).

The Z sources are characterized by three branches in an X-ray
color-color diagram: the horizontal, normal, and flaring branches--
NB, HB, and FB.  On the HB, these sources exhibit $\sim$ 13 - 55 Hz
QPO with a strong positive correlation between the QPO frequency and
the source intensity.  The strength of these QPO (fractional rms
variations up to $\sim$ 8\%) normally decreases with increasing source
intensity.  These QPO are accompanied by strong LFN whose strength
(fractional rms variations up to $\sim$ 8\%) has been reported to both
increase (Cyg X-2; Hasinger 1991) and decrease (GX 5-1; Lewin \etal
1992) with
increasing source intensity.  On the normal branch (NB), QPO of $\sim$
5 - 8 Hz are observed; the frequency of these oscillations is
approximately independent of source intensity, as is the strength of
the VLFN component.  The strength of both NB and HB QPO are strongly
dependent on X-ray energy.  Five of the Z sources
(Sco X-1, GX 340+0, GX 17+2, Cyg X-2, and GX 349+2) exhibit a FB
(Penninx \etal 1990b and references therein) whose characteristic
properties merge smoothly with those of the NB.  In the Flaring Branch
(FB) all but GX 340+0 show QPO with frequencies of $\sim$ 10 - 20 Hz
which are highly correlated with X-ray intensity (\hvdk; Penninx \etal
1990a,b).

Overall, the CCD and fast-timing characteristics of the Rapid Burster
bear no resemblance to those of either an atoll source or a Z-source. 

\subsection{The Persistent Emission}
We found that the hard and soft ratios of the PE observed Aug 28-29
1985 are not statistically correlated, while those observed Aug 30-31
1985 are statistically correlated.  The difference could be due to a
lack of spectral variability on Aug 28-29.

In the PE alone, specifically in the August 1985 observations, the
nature of the hard and soft colors looks similar to that seen in an
atoll source island state (``by eye''; no correlation statistics exist
for island states of atoll sources); however, spectral evolution of
the magnitude observed in either an atoll or Z-source could go
unnoticed in the August 1985 CCD, due to poor counting statistics.
The highly peaked QPO (\ie\ QPO power modelled by a function which has
a very highly positive derivative over some frequency range) are
unlike any timing characteristic in the PDS of atoll sources, most of
which do not exhibit peaked QPO (however, 1608-522 has exhibited QPO;
\cite{yoshida93}).  It is clear that the spectral hardness-correlated
VLFN in the PE is related to the humps.  It is also clear that the
characteristics of the peaked QPO (\%rms, width, centroid frequencies)
are independent of the hump; when the humps are excluded from the
analysis, these characteristics and their spectral correlations do not
change.  Very globally, the QPO centroid frequencies increase with
spectral hardness, the \%rms first decreases and then increases, and
the width of the QPO feature decreases (relative to the centroid
frequency) during spectrally harder PE (Tables~\ref{tab:peparams} \&
\ref{tab:perms}; Fig.~\ref{fig:bupermscomp}).

An alternative view of the timing properties in the $>$1 Hz range may
be to describe the QPO feature as HFN (such as is observed in atoll
sources in the Island state) which becomes increasingly peaked as the
energy spectrum hardens.  In this view, the QPO of the persistent
emission merges smoothly with the LFN, and thus the two could be of
common origin.

No excess \%rms in either the 10-100 or 20-60 Hz range (i.e. broad
HFN) was found in the PE with 2\sig~ upper limits of $\sim$ 8-10\%
(1-20 keV); these upper limits exceed the values measured for atoll
sources; it cannot therefore be excluded that a HFN atoll-like feature
is present in the PE in this frequency range.

The $\sim$ 0.040 Hz QPO are found only in the (spectrally hard) humps,
which sometimes occur following type II bursts.  No sub-Hz QPO have
been found in Z sources, and only one atoll source is known to exhibit
sub-Hz QPO (1608-522, \cite{yoshida93}; however, sub-Hz QPO has been
observed in black-hole candidates -- see, for a review, Tanaka \&
Lewin 1994 -- and from several pulsars, see
\cite{takeshima94} for a recent discussion).  For the
Rapid Burster, it is not unreasonable to assume that they are
transient phenomena which are associated with the poorly understood
accretion flow (see Lewin, Van Paradijs, \& Taam 1993, and references
therein).

\subsection{The Bursts}
The presence of QPO makes the RB during type II bursts very different
from atoll sources.  The behavior of the type II bursts of the Rapid
Burster somewhat resembles that of Z sources.  The CCDs of the long
bursts appear as resolved ``branches''.  Like in the NB of Z sources,
burst hardness is strongly correlated with source intensity
(\cite{kunieda84};
\cite{tan91};
\cite{lubin92a}; Figures~\ref{fig:HID85}a,b).  The strength of the
VLFN component in type II bursts is weakly correlated with spectral
hardness (Tables~\ref{tab:perms} \& ~\ref{tab:burstrms} ).  This
behavior is somewhat similar to the NB in Z sources.  Very globally,
the QPO strength is anticorrelated with source intensity, as is true
in both the HB and FB.

However, the Rapid Burster has striking differences from Z sources.
The QPO frequency and peak flux in the Rapid Burster are
anti-correlated (\cite{lubin91} and references therein); in Z sources,
the QPO frequency is positively correlated with intensity (in both the
HB and FB, where the QPO frequency is observed to change).  If the
``upper branch'' in the burst CCD of the 1985 observation were to be
interpreted as a Horizontal Branch, then the absence of LFN in Rapid
Burster bursts (see
\cite{stella88a}) demonstrates that the burst QPO are unlike HB
oscillations.  Further, the frequencies observed in the Rapid Burster
``Horizontal Branch'' (2-3 Hz) are very different from those in the HB
of Z sources (13-55 Hz) -- although we cannot rule out the existence
of QPO at these frequencies, since the upper limits in the Rapid
Burster are comparable to the measured strength in the Z sources' HB
(2-5\% rms).  There is also no evidence for broad HFN in the bursts,
with 2\sig~upper-limits comparable to the HFN strength observed in the
NB and HB of Z sources ($\sim$ 3- 4\%).  Nothing resembling the
Flaring Branch of a Z-source is observed.

\subsection{A Comparison with Cir X-1}
The Rapid Burster is not the only NS LMXB whose behavior is very
different from atoll or Z sources; Cir X-1 also does not always behave
clearly as one of these two classes, although at a certain orbital
phase, it has been observed to behave like an atoll source (see
\cite{oo94}, hereafter O94). The temporal variations of Cir X-1 are
very complicated (for a review, see e.g. \cite{dower82}). Cir X-1
emits type I X-ray bursts (Tennant, Fabian \& Shafer 1986a,b), as does the Rapid
Burster. QPO with centroid frequencies of 1.4 Hz and 5 - 18 Hz have
been observed (Tennant 1987; 1988a,b). They occur in different
spectral states.  The frequency of the 1.4 Hz QPO changed by less than
5\% as the source flux rose by a factor of 4. The 1.4 Hz QPO peak is
remarkably narrow with a FWHM between 0.03 and 0.06 Hz. The 5 - 18 Hz
QPO decreased in frequency with source intensity and appeared to
saturate at a value of about 5.7 Hz. The latter resembles the flaring
branch QPO in Z sources; however, the spectral hardness was strongly
anti-correlated with the source intensity as the 5 - 18 Hz QPO
occurred, and this is not observed in the flaring branch of most (but
not all) Z sources.

The combined spectral and timing behavior of Cir X-1 has been
investigated in depth by O94.  Almost all of the observed
spectral/timing behavior of Cir X-1 and the Rapid Burster are
dissimilar; however, there is some similarity between the behavior of
the Rapid Burster August 1985 PE and the zero-phase high-intensity
observation (H in O94) of Cir X-1.  During this observation, Cir X-1
had a very high intensity ($\sim$1500 {\it EXOSAT} c/s 1-20 keV).  As
the source spectrum hardened, the PDS evolved from VLFN only, to VLFN
and a LFN broad-band component with a cut-off frequency of $\sim$ 10
Hz, to broad LFN with peaked QPO near the cut-off frequency, to VLFN
with peaked QPO at $\sim$ 5-7 Hz.  The strength of the LFN and QPO
summed together ($\sim$5-7\%) remained largely constant as a function
of spectrum.

This is similar to the correlated spectral and timing behavior
observed from the Rapid Burster when it exhibited PE in August 1985.
During this observation there is some evidence for a VLFN component,
accompanied by a broad timing component modelled by a Lorentzian which
evolved into peaked QPO ($\nu_c \sim 3.0$ Hz) as the spectrum hardened
(Tables~\ref{tab:peparams} \& \ref{tab:nohumpparams};
Figs.~\ref{fig:pepds} \& \ref{fig:nohumppds}) .  However, if we were
to interpret this to mean that the Rapid Burster is in a similar
accretion state to Cir X-1, this would require that the Rapid Burster
is near the Eddington limit during the persistent emission, as the
behavior of Cir X-1 has been interpreted as being due to a near- or
super-Eddington accretion rate (\cite{vdk94}).  It seems unlikely that
the PE should correspond to such a high accretion rate, as the type II
bursts, which are cased by unstable accretion, have intensities which
are a factor of 10 greater than the persistent emission intensity,
with little or no observable LFN or VLFN, which is expected for
Eddington and super-Eddington accretion (and observed in Z-source
flaring branches, where super-Eddington accretion is thought to
occur).  Thus, while the changes of the PDS within with spectral
hardness seen in the PE of the RB are similar to those seen in Cir X-1
at very high intensities, we do not interpret the Rapid Burster PE
behavior as being due to $\sim$ Eddington accretion.

A working hypothesis which classifies atoll sources, Z sources,
black-hole candidates, and Cir X-1 has been suggested by Van der Klis
(1994).  He classifies the behavior of each in terms of differences in
magnetic field strength (absent in BHCs, absent or weak in atoll
sources, strong in Z sources) and accretion rate (which determines
which state the source is in --- low, high, or very high for BHCs;
island or banana for atoll sources; HB, NB, or FB for Z sources).
There is no obvious place to fit the Rapid Burster within this
framework, neither when we consider the spectral and timing behavior
of the PE or bursts separately, nor when they are considered together.

The spectral and timing behavior distinguishes the Rapid Burster from
other NS low-mass X-ray binaries, including Cir X-1.  The unknown
mechanisms which drive the variability of its accretion rate clearly
have significant effects on all facets of its appearance and behavior.

\pagebreak

\SectionN{\bf Acknowledgements}

The authors would like to thank Victoria Scheribel and Jennifer Carson
for aid in the preliminary data analysis. WHGL was supported by the
United States National Aeronautics and Space Administration (NASA)
under grant NAG8-700.  JvP acknowledges support from NATO through
grant RG 331/88.  This paper is supported in part by the Netherlands
Organization for Scientific Research (NWO) under grant PGS 78-277. LML
acknowledges support of NASA through grant NGT-51295. 

\pagebreak

\begin{table}[h]
\begin{center}
\scriptsize
\caption{Observational Data}
\label{tab:data}
\vspace{0.1cm}
\begin{tabular}{llcccc} \hline \hline
Date    &       
Period (UT)     &
OBC Modes$^{a}$ &
Energy Channels&
Time Resolution (ms)&
Transmission$^{b}$ (\%)\\ \hline
1983 Aug 15$^d$ & 00:57 - 09:59 & E4    & 8     &  93.75                & 14    \\
1984 Jul 17     & 00:50 - 06:17 & E5    & 32    &  312.5                & 17.5  \\
1985 Aug 28-29$^d$& 17:47 - 8:12  & E5  & 8, 32 &  312.5, 156.25,  93.75& 42    \\
                &               & I5    & 1     & 31.25                 &       \\
                &               & T3    & 1     &  3.9                  &       \\
1985 Aug 30-31$^d$& 18:28 - 7:24        & E5    & 64    & 1000                  &40.25, 23, 17.25$^c$   \\
                &               & I5    & 1     & 31.25                 &       \\
                &               & I7    & 1, 4  & 0.98, 3.9, 7.8        &       \\
                &               & T3    & 1     &  3.9                  &       \\
1985 Sept 10    & 12:10 - 18:22 & E5    & 64    & 500, 750              &43     \\
                &               & T3    & 1     &  1.95, 3.9            &       \\
1985 Sept 13    & 02:55 - 10:45 & E5    & 64    & 750                   &43     \\ 
                &               & T3    & 1     &  3.9                  &       \\
\hline
\multicolumn{6}{l}{$^a$Not all OBC modes are available during the whole
observation; ``E5''=HER5; ``I5''=HER5 (no energy resolution; ``T3''=HTR3} \\
\multicolumn{6}{l}{$^b$Includes aspect efficiency and correction to full array area} \\
\multicolumn{6}{l}{$^c$Detector halves swapped on and off source; see Sec.~\ref{sec:obs}} \\
\multicolumn{6}{l}{$^d$Persistent emission detected during these observations} \\
\end{tabular}
\end{center}
\end{table}

\begin{table}[h]
\begin{center}
\scriptsize
\caption{Counts Conversions}
\label{tab:mb}
\vspace{0.1cm}
\begin{tabular}{ccccc} \hline \hline
I7 range (keV) & to & E5 range (keV)    &        slope ($\pm$; E5 Counts/I7 Counts)     & intercept ($\pm$; E5 Counts) \\ \hline \hline
\multicolumn{5}{c}{Persistent Emission} \\ \hline
1.2-3.4& &0.9-2.6 & 0.55 $\pm$ 0.02 & 0.62 $\pm$ 0.53 \\
3.7-5.5&&2.9-4.7 & 1.02 $\pm$ 0.017 & 2.36 $\pm$ 0.44 \\
5.8-10.3&&5.0-6.9& 0.92 $ \pm$ 0.024 & 1.91 $\pm$ 0.43 \\ \hline
\multicolumn{5}{c}{Bursts} \\ \hline 
1.2-3.4&&0.9-2.6 & 0.48 $\pm$ 0.009 & 3.6 $\pm$ 1.5 \\
3.7-5.5&&2.9-4.7 & 0.98 $\pm$ 0.007 & 6.5 $\pm$ 1.5 \\
5.8-10.3&&5.0-6.9& 1.02 $\pm$ 0.008 & 4.4 $\pm $ 1.1 \\
\hline
\end{tabular}
\end{center}
\end{table}

\begin{table}[h]
\begin{center}
\scriptsize
\caption{August 1985 Persistent Emission PDS Fit Parameters }
\label{tab:peparams}
\vspace{0.1cm}
\begin{tabular}{lcccccccc} \hline \hline

Hard Color & $\chi^2$ (dof)     & $\alpha$                      & $\nu_{0.04 Hz}$       & $\Gamma_{0.04 Hz} $   & $\nu_1$                       & $\Gamma_1 $           & $\nu_2$               & $\Gamma_2 $   \\ \hline
0.66-0.74       &78 (50)        & -                             & -                     & -                          & $0.075^{+0.44}_{-1.1}$   & $2.9^{+0.7}_{-0.6}$   & -                     & -                        \\
0.74-0.78       &48 (50)        & $2.0 \pm 0.3$                 & -                     & -                          & $-1.1^{+1.6}_{-2.5}$     & $10^{+1.4}_{-1.2}$    & -                     & -                        \\
0.78-0.82       &83 (50)        & $1.0 \pm 0.1$                 & -                     & -                          & $2.8 \pm 0.1$            & $1.6 \pm 0.2  $       & -                     & -                        \\
0.82-0.86       &62 (49)        & $2.1 \pm 0.2$                 &$0.041^a$              & $0.0011^a$                 & $2.1 \pm 0.2$            & $5.2^{+0.6}_{-0.5}$   & -                     & -                        \\
0.86-0.90       &64 (44)        & $2.3 \pm 0.3$                 &$0.040 \pm 0.03$       & $0.0016^{+0.011}_{-0.007}$ & $2.49 \pm 0.04$          & $0.56^{+0.09}_{-0.07}$&$3.57 \pm 0.05$        & $0.7 \pm 0.1$         \\
0.90-1.02       &58 (44)        & $1.28 \pm 0.04$               &$0.0427 \pm 0.0007$    & $0.0010 \pm 0.0004$        & $2.88 \pm 0.02$          & $0.33^{+0.08}_{-0.11}$&$3.97 \pm 0.05$        & $1.2 \pm 0.06$                \\
\hline
\multicolumn{9}{l}{$^a$ Values were fixed} \\
\end{tabular}
\end{center}
\end{table}

\begin{table}[h]
\begin{center}
\scriptsize
\caption{August 1985   Persistent Emission \% rms Values from Parametric Fits}
\label{tab:perms}
\vspace{0.1cm}
\begin{tabular}{lccccc} \hline \hline
Hard Color      &       VLFN$^a$        &       QPO ($\sim$ 0.04 Hz) &  QPO~($\nu_1$)   &QPO~($\nu_2$)  &QPO($\nu_1 + \nu_2$)    \\
                &       \% rms          & \% rms                      & \% rms          &       \% rms          &       \% rms          \\ \hline
0.66-0.74       &       $ \le 2.9^b    $& -             &$14 \pm 2   $          & -                     &$14 \pm  2$            \\
0.74-0.78       &       $5.6 \pm 0.8   $& -             &$16.6^{+1.0}_{-1.5}$   & -                     &$16.6^{+1.0}_{-1.5}$           \\
0.78-0.82       &       $7.1 \pm 0.4   $& -             &$12.0^{+0.6}_{-0.8}$   & -                     &$12.0^{+0.6}_{-0.8}$           \\
0.82-0.86       &       $7.4 \pm 0.6   $&$\le 2.5^b$    &$17.4 \pm 0.7$         & -                     &$17.4 \pm 0.7$         \\
0.86-0.90       &       $10.2^{1.5}_{-2.0}$&$6.9 \pm 1.3$&$11.4 \pm 0.9$        &$12.7 \pm 0.9$         &$17.1 \pm 1.9$          \\
0.90-1.02       &       $17.5 \pm 0.7 $ &$\le 84^{b,c}$  &$14.0^{+0.9}_{-1.7}$   &$22.2^{+0.4}_{-1.1}$   &$26.2^{+1.7}_{-3.4}$         \\
\hline
\multicolumn{6}{l}{$^a$ Integrated from 0.0078-1.0 Hz} \\
\multicolumn{6}{l}{$^b$ 2\sig~upper limits} \\
\multicolumn{6}{l}{$^c$ Poorly constrained due to poor frequency resolution} \\
\end{tabular}
\end{center}
\end{table}

\begin{table}[h]
\begin{center}
\scriptsize
\caption{August 1985 Persistent Emission PDS Fit Parameters (Hump Excluded)}
\label{tab:nohumpparams}
\vspace{0.1cm}
\begin{tabular}{lcccccc} \hline \hline

Hard Color & $\chi^2$ (dof)     & $\alpha$                              & $\nu$                 & $\Gamma $     & $\nu$                 & $\Gamma $     \\ \hline
0.66-0.74       &70 (52)        & $2.00^a$                      & $0.6 \pm 0.2$         & $1.6^{+0.4}_{-0.3}$   &       -               &       -       \\
0.74-0.78       &60 (52)        & $2.6^{+0.8}_{-0.6}$           & $0.1^{+0.4}_{-0.5}$   & $4.73^{+0.9}_{-0.6}$& -               &       -       \\
0.78-0.82       &77 (52)        & -                             & $2.7 \pm 0.1$         & $1.8 \pm 0.2$         &       -               &       -       \\
0.82-0.86       &53 (52)        & $2.5 \pm 0.1$                 & $1.8 \pm 0.2$         & $4.5 \pm 0.4$         &       -               &       -       \\
0.86-0.90       &110 (49)       & $2.8 \pm 0.1$                 & $2.45 \pm 0.05$       & $0.48 \pm 0.07$       &       $3.43 \pm 0.06$ & $0.6 \pm 0.2$\\
0.90-1.02       &84 (52)        & $1.77\pm 0.02$                & $3.42 \pm 0.03$       & $0.53 \pm 0.09$       &       -               &       -       \\
\hline
\multicolumn{7}{l}{$^a$ Value fixed} \\
\end{tabular}
\end{center}
\end{table}

\begin{table}[h]
\begin{center}
\scriptsize
\caption{ August 1985  Persistent Emission \% rms Values from Parametric Fits (Hump Excluded)}
\label{tab:nohumpperms}
\vspace{0.1cm}
\begin{tabular}{lcc} \hline \hline
Hard Color &    VLFN$^a$                &       QPO~($\nu_1 +\nu_2$)    \\
        &       \% rms                  & \% rms (0.01-100 Hz)          \\ \hline
0.66-0.74       &       $ \le 2.2^b    $        &$      12.7 \pm 2.7$                   \\
0.74-0.78       &       $ \le 4.4^b     $       &$      15.9^{+1.8}_{-2.3}$                     \\
0.78-0.82       &       $ 2.8^{+0.9}_{-0.7}   $ &$      15.3^{+2.8}_{-3.1}$                     \\
0.82-0.86       &       $ 4.4^{+0.7}_{-0.8} $   &$      16.4^{+1.2}_{-1.0}$                     \\
0.86-0.90       &       $ 7.5^{+1.4}_{-1.8}$    &$      16.7^{+1.3}_{-1.5}$      \\
0.90-1.02       &       $ 3.5^{+0.6}_{-0.6}$    &$      20.4^{+0.9}_{-0.8}$             \\
\hline
\multicolumn{3}{l}{$^a$ Integrated from 0.0078-1.0 Hz} \\
\multicolumn{3}{l}{$^b$ 2\sig~upper limits} \\
\end{tabular}
\end{center}
\end{table}

\begin{table}[h]
\begin{center}
\caption{Weighted Average Type II Burst Hardness Ratios}
\label{tab:avghardratios}
\vspace{0.1cm}
\begin{tabular}{lcc} \hline \hline
Date            &       3-5 keV/ 1-3 keV        &       5-10 keV/ 3-5 keV       \\ \hline
1984 Jul 17     &       $3.09 \pm 0.04$         &       $0.998 \pm 0.007$       \\
1985 Aug 28-29  &       $2.93 \pm 0.03$         &       $0.94 \pm 0.01$         \\
1985 Aug 30-31  &       $2.98 \pm 0.01$         &       $0.99 \pm 0.01$         \\
1985 Sep 10     &       $2.88 \pm 0.01$         &       $0.943 \pm 0.003$       \\
1985 Sep 13     &       $2.77 \pm 0.01$         &       $0.892 \pm 0.003$       \\
\hline
\end{tabular}
\end{center}
\end{table}

\begin{table}[h]
\begin{center}
\caption{August 1985 Burst Fit Parameters }
\label{tab:burstparams}
\vspace{0.1cm}
\begin{tabular}{lcccccc} \hline \hline
Hard Color      &$\chi^2$ (dof) &$\alpha$       &$\nu_1$                &$\Gamma_1 $            &$\nu_2$&$\Gamma_2 $  \\ \hline
0.74-0.80       &60 (49)&       2.00$^a$        &$4.71 \pm 0.04$        & $0.89 \pm 0.08$       & -                     & -  \\
0.80-0.86       &50 (49)&       2.00$^a$        &$4.87 \pm 0.02$        & $0.67 \pm 0.04$       & -                     & -  \\
0.86-0.92       &57 (49)&$1.8 \pm 0.3$          &$4.83 \pm 0.03$        & $0.85 \pm 0.07$       & -                     & -  \\
0.92-0.98       &32 (49)&$1.4 \pm 0.2$          &$4.45 \pm 0.14$        & $1.7 \pm 0.3$         & -                     & -  \\
0.98-1.04       &53 (49)&$1.8 \pm 0.2$         &$2.20 \pm 0.04$        & $0.43^{+0.16}_{-0.08}$& -                     & -  \\
1.04-1.12       &40 (46)&$2.2 \pm 0.3$         &$2.03 \pm 0.05$        & $0.08 \pm 0.02$       & $2.36 \pm 0.03$       & $0.36 \pm 0.06$ \\ \hline
\multicolumn{7}{l}{$^a$ Fixed} \\
\end{tabular}
\end{center}
\end{table}

\begin{table}[h]
\begin{center}
\caption{August 1985 Burst \% rms Values from Parametric Fits }
\label{tab:burstrms}
\vspace{0.1cm}
\begin{tabular}{lcc} \hline \hline
Hardness &              VLFN$^a$        &       QPO~($\nu_1+ \nu_2$)    \\
         &              \% rms  &       \% rms                  \\\hline
0.74-0.80       &       $\le 2.5^b$     &       $25.6 \pm 0.9$  \\
0.80-0.86       &       $\le 2.4^b$     &       $21.2^{+1.3}_{-0.2}     $\\
0.86-0.92       &       $2.0 \pm 0.6$   &       $14.0^{+1.3}_{-0.1}$\\
0.92-0.98       &       $2.2 \pm 0.3$   &       $ 5.8 \pm 0.4$  \\
0.98-1.04       &       $2.3 \pm 0.3$   &       $ 4.6 \pm 0.4$  \\
1.04-1.12       &       $2.3 \pm 0.3$   &       $10.7^{+7.6}_{-1.1} $\\
\hline
\multicolumn{3}{l}{$^a$ Integrated from 0.06-1.0 Hz} \\
\multicolumn{3}{l}{$^b$ Upper limits are 2\sig~confidence} \\
\multicolumn{3}{l}{$^c$ Sum of two fitted Lorentzians} \\
\end{tabular}
\end{center}
\end{table}

\begin{table}[h]
\begin{center}
\footnotesize
\caption{September 1985 Bursts PDS Fit Parameters }
\label{tab:sept85burstfitparams}
\vspace{0.1cm}
\begin{tabular}{lccccccc} \hline \hline
                        & \multicolumn{2}{c}{Primary$^b$ Fit}   & & \multicolumn{4}{c}{Residual Fit}                                                    \\ \cline{2-3} \cline{5-8}
Peak Count Rate$^a$     & $\chi^2$ (dof)&  VLFN             & & $\chi^2$ (dof)   &VLFN           &    \multicolumn{2}{c}{QPO}                  \\ \cline{7-8}
        (counts/sec)    &               & $\alpha$              & &                  &$\alpha$            & $\nu_c$ (Hz)           &$\Gamma$ (Hz)       \\ \hline
300  $< I_{peak} <$ 1000& 18.3 (15)     & $3.05 \pm 0.02$       & & 248 (21)         &$ 0.8^c$            & $3.97 \pm 0.2$          &$5 ^{+0.6}_{-0.2}$      \\
1000 $< I_{peak} <$ 1100& 56.2 (15)     & $3.13 \pm 0.02$       & & 192 (21)         &$0.7^{+0.3}_{-0.1}$& $5.08 \pm 0.09$         &$2.52 \pm 0.2$      \\
1100 $< I_{peak} <$ 1500& 30.3 (15)     & $3.08 \pm 0.02$       & & 84 (21)          &$0.73 \pm 0.05$    & $5.82 \pm 0.03$         &$0.96 \pm 0.15$     \\
All                     & 62.6 (15)     & $3.11 \pm 0.01$       & & 363 (21)         &$0.91 \pm 0.07$   & $5.19 \pm 0.07$          &$2.7 \pm 0.2$      \\
\hline 
\multicolumn{8}{l}{$^a$ Corrected to full detector area and for aspect, but not deadtime} \\
\multicolumn{8}{l}{$^b$ Power law fit above 0.6 Hz} \\
\multicolumn{8}{l}{$^c$ Value Fixed} \\
\end{tabular}
\end{center}
\end{table}

\begin{table}[h]
\begin{center}
\scriptsize
\caption{September 1985 Bursts Parametric Fit Results }
\label{tab:sept85burstrms}
\vspace{0.1cm}
\begin{tabular}{lcccc} \hline \hline
Peak Count Rate$^a$     & Primary$^b$ VLFN$^c$  &      Residual VLFN$^c$ & Primary + Residual VLFN$^c$  &       QPO             \\
 (counts/sec)           &               \% rms  &               \% rms  &               \% rms          &       \% rms          \\\hline
 300  $< I_{peak} <$ 1000&       $52 \pm 2$     &       $ <6.7^d$  &           $53 \pm 2$          &       $21 \pm 1.0$  \\
 1000 $< I_{peak} <$ 1100&       $48 \pm 2$     &       $4.4 \pm 0.4$  &       $48 \pm 2$        &         $15.5 \pm  0.7$  \\
 1100 $< I_{peak} <$ 1500&       $52 \pm 2$     &       $6.9 \pm 0.5$  &       $52 \pm 2$          &       $7.6 \pm 0.5$  \\
 All                     &       $50 \pm 2$     &       $7.5 \pm 0.5$  &       $51 \pm 2$          &       $13.9 \pm 0.7$  \\
\hline
\multicolumn{5}{l}{$^a$ Corrected to full detector area and for aspect, but not deadtime} \\
\multicolumn{5}{l}{$^b$ Power law fit above 0.6 Hz} \\
\multicolumn{5}{l}{$^c$ Integrated from 0.125-1.0 Hz} \\
\multicolumn{5}{l}{$^d$ Upper limits are 2$\sigma$} \\
\end{tabular}
\end{center}
\end{table}

\clearpage
\pagebreak

\begin{figure}
\caption{ \label{fig:83pehidccd}
HIDs and CCD for the persistent emission observed 15 Aug 1983 (type I
bursts were excluded).  The count rates have been background
subtracted, aspect corrected, and normalized to the full array.  No
dead time corrections (which are $\simeq$ 1\%) have been performed.
Because of a different OBC mode, the energy ranges used in the
analysis of this observation were different from those used in the
other observations, which complicates a direct comparison between
these HIDs and CCDs and those of other observations. Each point in all
three panels represents 600 seconds of integration.  a) HID, 2-17 keV
count rate vs. soft color (5.7--9.3 keV/2.0--5.7 keV). b) HID, 2-17
keV count rate vs hard color (9.3--17.0 keV/5.7--9.3 keV). c)
Color-color diagram. soft color vs. hard color. }
\end{figure}


\begin{figure}
\caption{ \label{fig:85pehidccd}
HIDs and CCD for the persistent emission observed 28-29 Aug 1985 and
30-31 Aug 1985.  The observation date is indicated at the top of each
panel. The count rates have been background subtracted, dead time
corrected, aspect corrected, and normalized to the full array.  An
average 1-sigma error bar is included in each panel. a) 28-29 Aug 1985
HID, 1-20 keV count rate vs. hard color. b) 28-29 Aug 1985 HID, 1-20
keV count rate vs. soft color.  c) 28-29 Aug 1985 CCD, soft color vs.
hard color. d) 30-31 Aug 1985 HID, 1-20 keV count rate vs.  hard
color. e) 30-31 Aug 1985 HID, 1-20 keV count rate vs. soft color. f)
30-31 Aug 1985 CCD, soft color vs.  hard color.  }
\end{figure}

\begin{figure}
\caption{ \label{fig:85intdist}
Distribution of the average (1 - 20 keV) intensities (a) and the soft
(3 - 5 keV/1 - 3 keV) (b) and hard color (5 - 10 keV/3 - 5 keV) (c) of
the persistent emission intervals from the two Aug 1985 data sets.
The integration time for each point was 100 sec.  All count rates
corrected for aspect, dead time (see Stella \& Andrews 1985; Eq. 3.10
of Van der Klis 1989), and normalized to the full array area. The persistent
emission observed during the August 30-31 observation is clearly {\it
harder} than that on August 28-29 (panel c) in the hard color, while
it is {\it softer} in the soft color.}
\end{figure}

\begin{figure}
\caption{ \label{fig:barrpds}
PDS of PE observed during the 1983 observations.  The included energy
range was 2-12 keV.  The 12 FFTs were each 1536 sec long, with a
Nyquist Frequency of 5.6 Hz.  The FFTs were summed into a single PDS,
which was logarithmically rebinned. Panels show the averaged PDS of
these hardness bins, and the averaged PDS of all FFTs together.  There
is no obvious VLFN in any of these PDS.  2\sig\ upper limits in VLFN
for the PDS of all 215 FFTs in the frequency range 0.005-1.0 Hz is
2.0\%. }
\end{figure}

\begin{figure}
\caption{ \label{fig:pepds}
Averaged PDS and best fit model of the persistent emission during the
1985 August 28-29 \& 30-31 observations for six ranges of the hard
color (5 - 10 keV/3 - 5 keV).  The range of the hard color and the
number of FFTs in this range are indicated in the upper right of each
panel.  The best fit Poisson level has been subtracted, and the PDS
has been logarithmically rebinned.  The filled squares with arrows
pointing downwards indicate 2$\sigma$ upper limits. The line is the
best fit model to all data points (including those consistent with
zero power). The frequency of the QPO peak increases with increasing
hardness.  The $\sim$ 40 mHz oscillations are only present in the
portions of persistent emission which are relatively hard (panels
e,f).  }
\end{figure}    

\begin{figure}
\caption{ \label{fig:nohumppds}
Averaged PDS and best fit model of the persistent emission during the
1985 August 28-29 \& 30-31 observations for six ranges of the hard
color (5 - 10 keV/3 - 5 keV), excluding the first 256 sec of PE (\ie\
the ``hump'').  The range of the hard color and the number of FFTs in
this range are indicated in the upper right of each panel.  The best
fit Poisson level has been subtracted, and the PDS has been
logarithmically rebinned.  The filled squares with arrows pointing
downwards indicate 2$\sigma$ upper limits. The line is the best fit
model to all data points (including those consistent with zero
power). }
\end{figure}

\begin{figure}
\caption{ \label{fig:hidccdcomp}
HIDs and CCDs of type II bursts observed in 5 of the 6 observation
periods of the Rapid Burster with the {\it EXOSAT} ME instrument. This
figure shows all type II bursts which were observed with $\ge$4 energy
channels by the {\it EXOSAT} ME instrument.  The 1-20 keV peak count
rates (in counts/sec) are averaged over $\sim$ 1 sec, are corrected
for dead time (see Stella \& Andrews 1985) and aspect, and are
normalized to the full array.  Each panel contains an average error
bar in the upper right-hand corner.  The observation date is indicated
at the top of each column of panels.  Note that 1 count/s $\simeq$ 1.6
$\times$ $10^{-11}$ ergs cm$^{-2}$ s$^{-1}$; see Tan
\etal (1991). }
\end{figure}

\begin{figure}
\caption{ \label{fig:burstcomp}
Composite HIDs and CCD of type II bursts observed by EXOSAT.  The
legend in the figure indicates the observation period for each
data point.  All count rates are corrected for deadtime, aspect, and
normalized to the full EXOSAT ME array area.  The July 1984 and
September 1985 data points represent the average hardnesses of the
($\ge$10) bursts in a peak count rate range indicated by the horizontal
bars.  The 1\sig\ errors in mean hardness for the July, 1984 and
September, 1985 data are indicated in each point, assuming the burst
hardnesses in each peak count rate range are normally distributed.  An
average 1\sig~error bar is included on one point for the August 28-29
and 30-31, 1985 data.  a) Soft color vs.  burst peak count rate (c/s).
b) Hard color vs. burst peak count rate (c/s). c) Hard color vs. soft
color.  }
\end{figure}

\begin{figure}
\caption{ \label{fig:burstpds}
Averaged PDS, logarithmically rebinned, of the type II bursts during
the 1985 August 28-29 \& 30-31 observations for six ranges of the hard
color (5 - 10 keV/3 - 5 keV).  The solid lines indicate the best fit
models (Table~\protect{\ref{tab:burstparams}}).  The number of FFTs and the
range of hardness of each averaged PDS are indicated in the upper left
of each panel.  The QPO frequency decreases from $\sim$ 5 Hz to $\sim$
2 Hz with increasing hard color while the \% rms in the QPO at first
decreases, and then increases with increasing hard color.}
\end{figure}    

\begin{figure}
\caption{ \label{fig:notcesme}
  Comparison of signal power with Poisson power.  (a)~PDS from 91
  burst FFTs (from Sept. 10-13 1985) using T3 data.  The Poisson level
  was not subtracted from the power. (b)~The ratio of \sig\ at each
  frequency to the power at each frequency. (c)~The uncertainty in the
  mean power (\sig ) at each frequency, which is found from the scatter
  of the 91 different powers from the 91 different FFTs at each
  frequency. The divergence of the ratio of \sig /Power from $1/({\rm NFFTs})^1/2$ at
  frequencies below 0.5 Hz demonstrates that the power at these
  frequencies is largely deterministic.  }
\end{figure}

\begin{figure}
\caption{  \label{fig:985totpds} 
  PDS of 312 type II bursts observed September 10-13, 1985 using OBC
  mode T3. Solid lines indicate the sum of the Primary and Residual
  models (see text).  The bursts were separated by peak intensity,
  which was measured in $\sim$ 1 sec bins, corrected to the full array
  area for aspect, but not for deadtime.  Error bars are 1$\sigma$,
  while the upper limits are 2\sig.  Solid symbols with arrows
  pointing downwards indicate 2\sig~upper limits.  {\bf (a)}~Averaged
  PDS of FFTs from 100 bursts with peak intensity below 1000 c/s.
  {\bf (b)}~Averaged PDS of FFTs from 116 bursts with peak intensity
  between 1000 and 1100 c/s.  {\bf (c)}~Averaged PDS of FFTs from 96
  bursts with peak intensity above 1100 c/s. {\bf (d)}~Averaged PDS of
  all 312 bursts.}
\end{figure}

\begin{figure}
\caption{ 
\label{fig:HID85}
Hardness-intensity diagrams (HIDs) of the bursts and the persistent
emission for the soft ratio (3 - 5 keV/1 - 3 keV) [(a)] and the hard
ratio (5 - 10 keV/3 - 5 keV) [(b)].  The data points from the August
28-29 and August 30-31, 1985 observations are indicated with closed
and open symbols, respectively.  For the burst data, each point
represents a single burst.  For the persistent emission data, each
point represents an integration of 100 sec.  For the bursts, 1 count/s
$\simeq$ 1.6 $\times$ $10^{-11}$ ergs cm$^{-2}$ s$^{-1}$; for the
persistent emission, 1 count/s $\simeq$ 7.8 $\times$ $10^{-12}$ ergs
cm$^{-2}$ s$^{-1}$ ( see Tan \etal 1991).  All count rates have been
normalized to the full array area.  Corrections for collimator
transmission (aspect), background, and dead time (see Stella \&
Andrews 1985) have also been made. A sample error bar is shown, offset
from the data, separately for the persistent emission and the bursts.
The previously reported correlation between burst peak counting rate
and hardness (or equivalently temperature) is apparent. }
\end{figure}    

\begin{figure}
\caption{ \label{fig:85CCDcomp}
The composite CCD for bursts and persistent emission during the August
28-29 and August 30-31, 1985 observations.  The bursts observed on
August 28-29 and August 30-31 are indicated by closed and open
squares, respectively.  For bursts, each data point represents a
single burst.  The persistent emission observed on August 28-29 and
August 30-31 are indicated by closed and open triangles, respectively.
Each data point represents an integration of 100 sec.  A sample error
bar for the burst and the persistent emission data is shown on one
data point.}
\end{figure}    

\begin{figure}
\caption{ \label{fig:burstpe}
a) The soft color (3 - 5 keV/1 - 3 keV)  and b) the hard color (5
- 10 keV/3 - 5 keV) versus burst number.   All bursts are numbered
sequentially in accordance with the numbering of Stella \etal (1988a)
and Lubin \etal (1992b).  Solid and open points indicate the hardness
of the burst and the {\it following} persistent emission interval,
respectively.  (Some of the bursts and persistent emission data
could not be included due to telemetry losses and the contamination of
the nearby source 1728-34.)  A global correlation between the hardness
of the bursts and the persistent emission is most evident in the hard
ratio (b).}
\end{figure}    

\begin{figure}
\caption{ \label{fig:hardbpe}
The hardness ratios of the bursts versus those of the {\it following}
persistent emission interval; the soft color (3 - 5 keV/1 - 3 keV) for
the August 28-29 (a) and 30-31 (b) observation and the hard color (5 -
10 keV/3 - 5 keV) for the August 28-29 (c) and 30-31 (d) observations.
There are statistical correlations in all four comparisons, with a
probability of being generated from uncorrelated data of: (a) 0.13\%,
(b) 0.06\%, (c) 0.04\%, and (d) 0.01\% (see text).}
\end{figure}    

\begin{figure}
\caption{ \label{fig:bupermscomp}
Comparison of the \% rms values for bursts and persistent emission
(PE) in the 1985 August 28-29 and 30-31 observations with the
PE ``humps'' excluded. Error bars are 1\sig~for the \%rms and indicate
the hardness range for the hardness ratio.  Upper-limits are
2\sig. {\bf (a)} \%rms values of the 2-5 Hz QPO as a function of
hardness interval for PE and bursts. The trends of \%rms for the PE
and bursts are oppositely directed. {\bf (b)}
\%rms of the VLFN in the PE and bursts. }
\end{figure}

\begin{figure}
\caption{ \label{fig:z}
HID of bursts observed during August 1985 (hard color vs. 1-20 keV
count rate).  Aug 28-29 bursts are indicated by a circle; Aug 30-31
bursts are indicated by a square.  Bursts which had a QPO peak between
2-7 Hz which could be identified by eye have the approximate centroid
frequency indicated at each point.  Dots indicate that no QPO was
seen.  No mark indicates that a complete PDS is not available.  The
broken lines divide the HID into groups of bursts, separating roughly
along the separate ``branches'' (arbitrarily drawn), labelled A, B, C, D,
\& E as indicated. An average error bar is included in the bottom
right corner.}
\end{figure}

\begin{figure}
\caption{ \label{fig:zpds}
  PDS of August 1985 bursts grouped as indicated in Figure
  \protect{\ref{fig:z}}.  The group is indicated in the upper left
  corner. The PDS panels are ordered on the figure roughly as the
  corresponding position on the HID in Figure \protect{\ref{fig:z}}.
  Error bars for the power are included on each point.  The Poisson
  level has not been subtracted off, and the data have been
  logarithmically rebinned.  }
\end{figure}


\clearpage
\pagebreak

\clearpage
\pagestyle{empty}
\begin{figure}
\PSbox{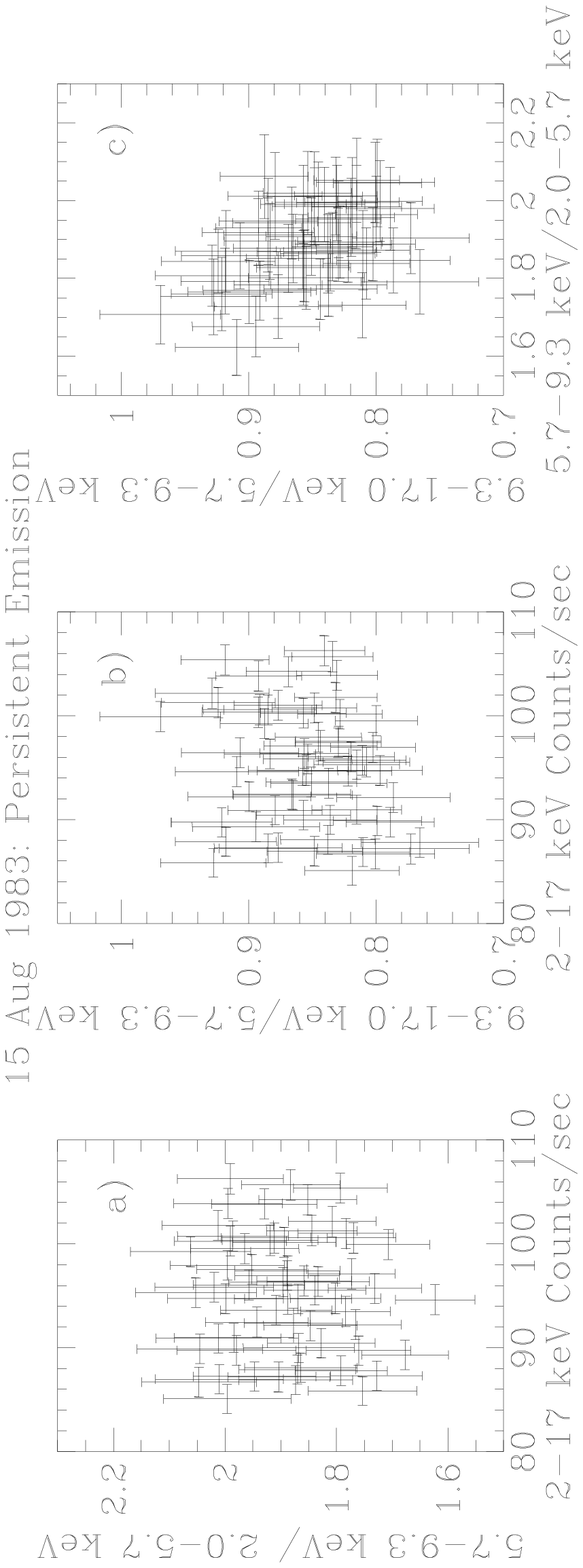 hoffset=-60 voffset=-56}{14.7cm}{21.5cm}
\FigNum{\ref{fig:83pehidccd}}
\end{figure}


\clearpage
\pagestyle{empty}
\begin{figure}
\PSbox{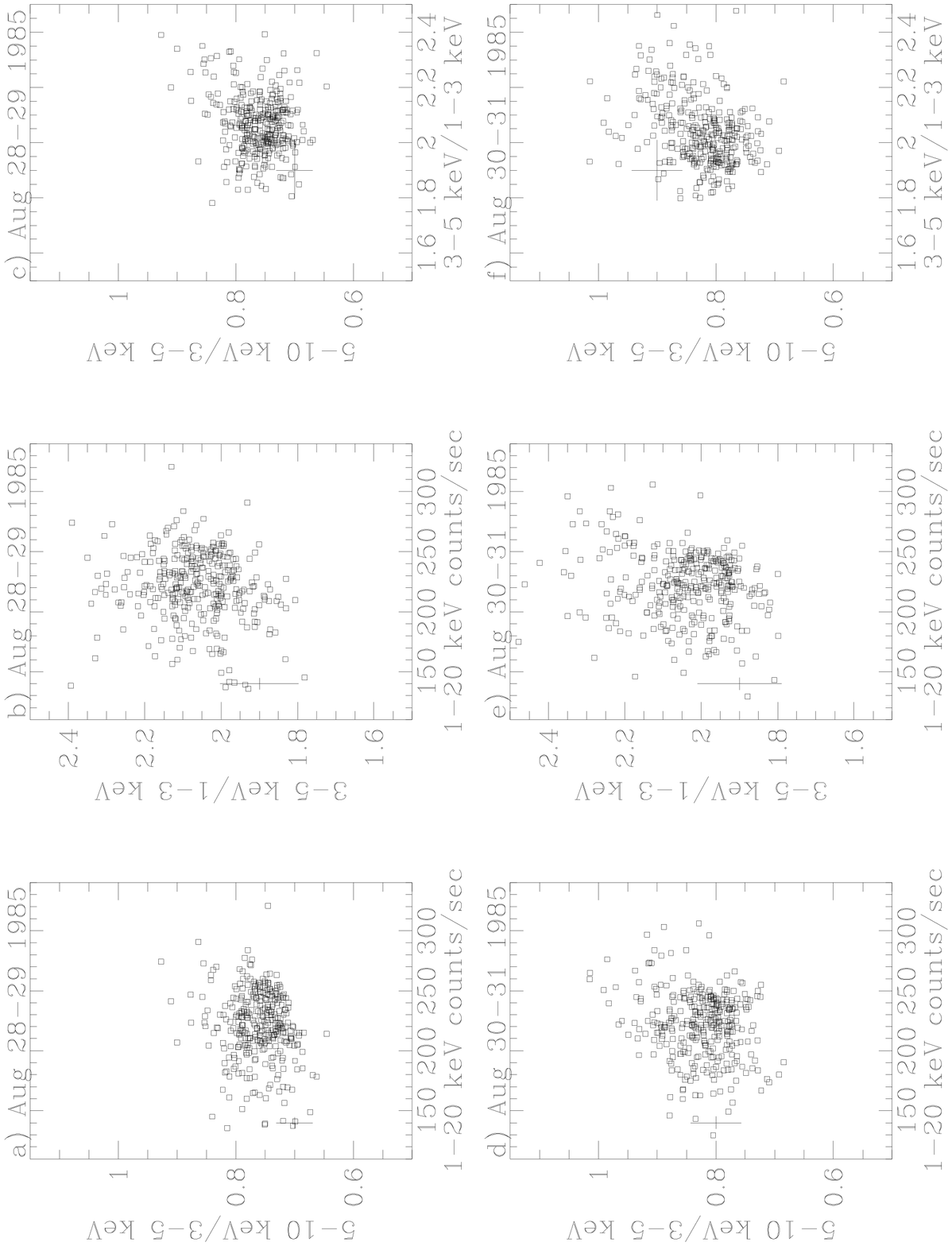 hoffset=-80 voffset=-56}{14.7cm}{21.5cm}
\FigNum{\ref{fig:85pehidccd}}
\end{figure}

\clearpage
\pagestyle{empty}
\begin{figure}
\PSbox{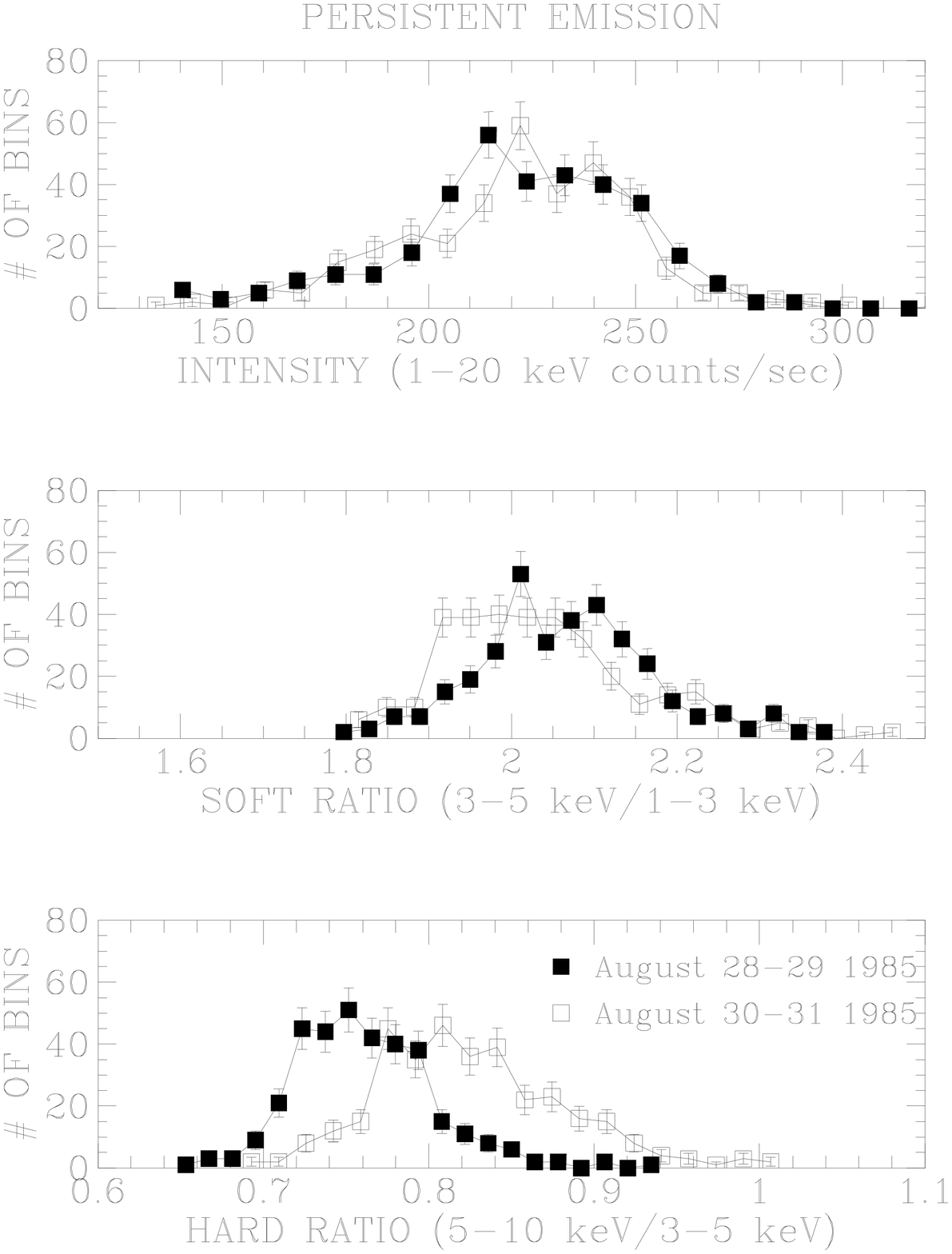 hoffset=-60 voffset=-56}{14.7cm}{21.5cm}
\FigNum{\ref{fig:85intdist}}
\end{figure}

\clearpage
\pagestyle{empty}
\begin{figure}
\PSbox{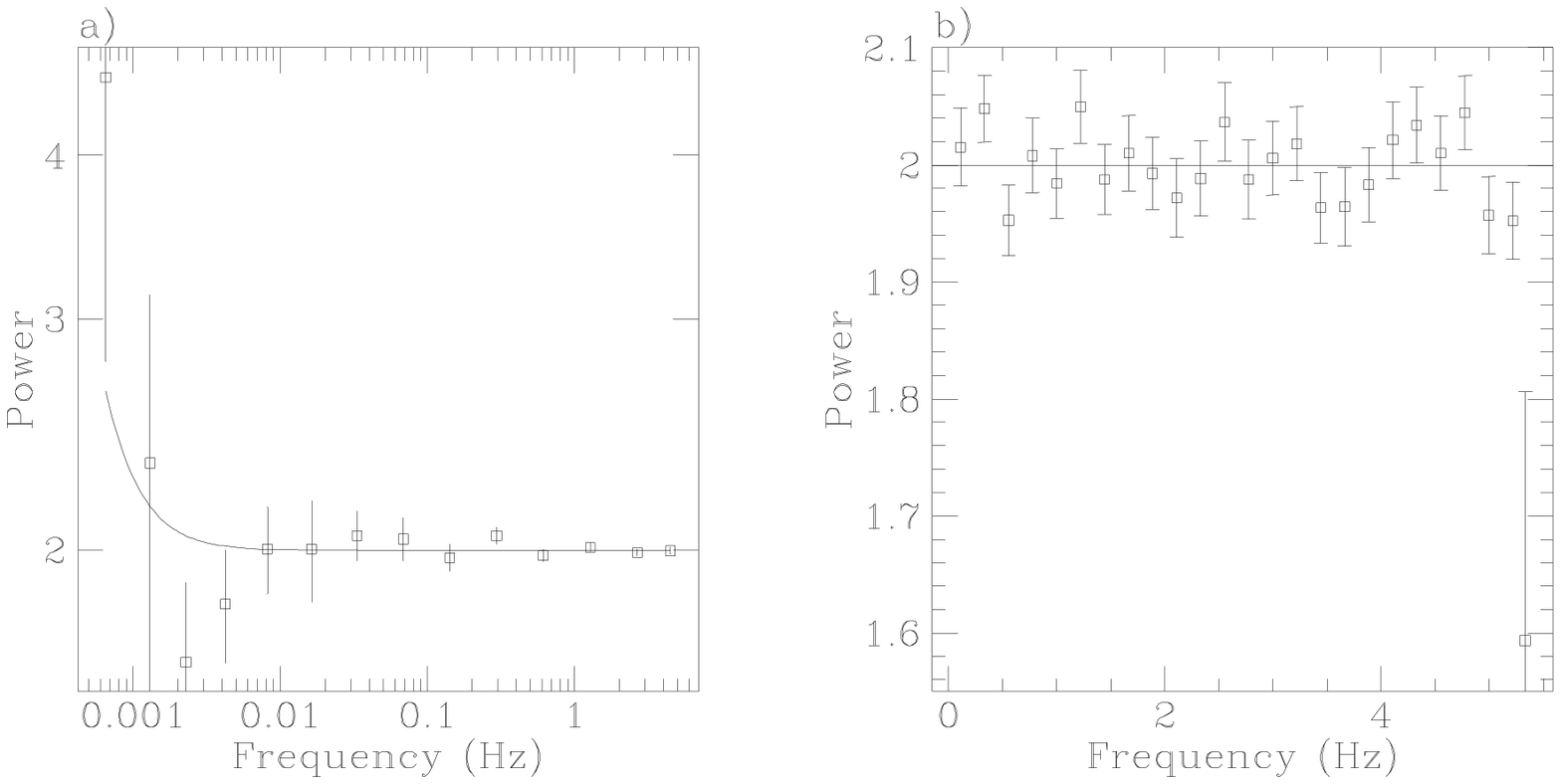 hoffset=-80 voffset=-56}{14.7cm}{21.5cm}
\FigNum{\ref{fig:barrpds}}
\end{figure}

\clearpage
\pagestyle{empty}
\begin{figure}
\PSbox{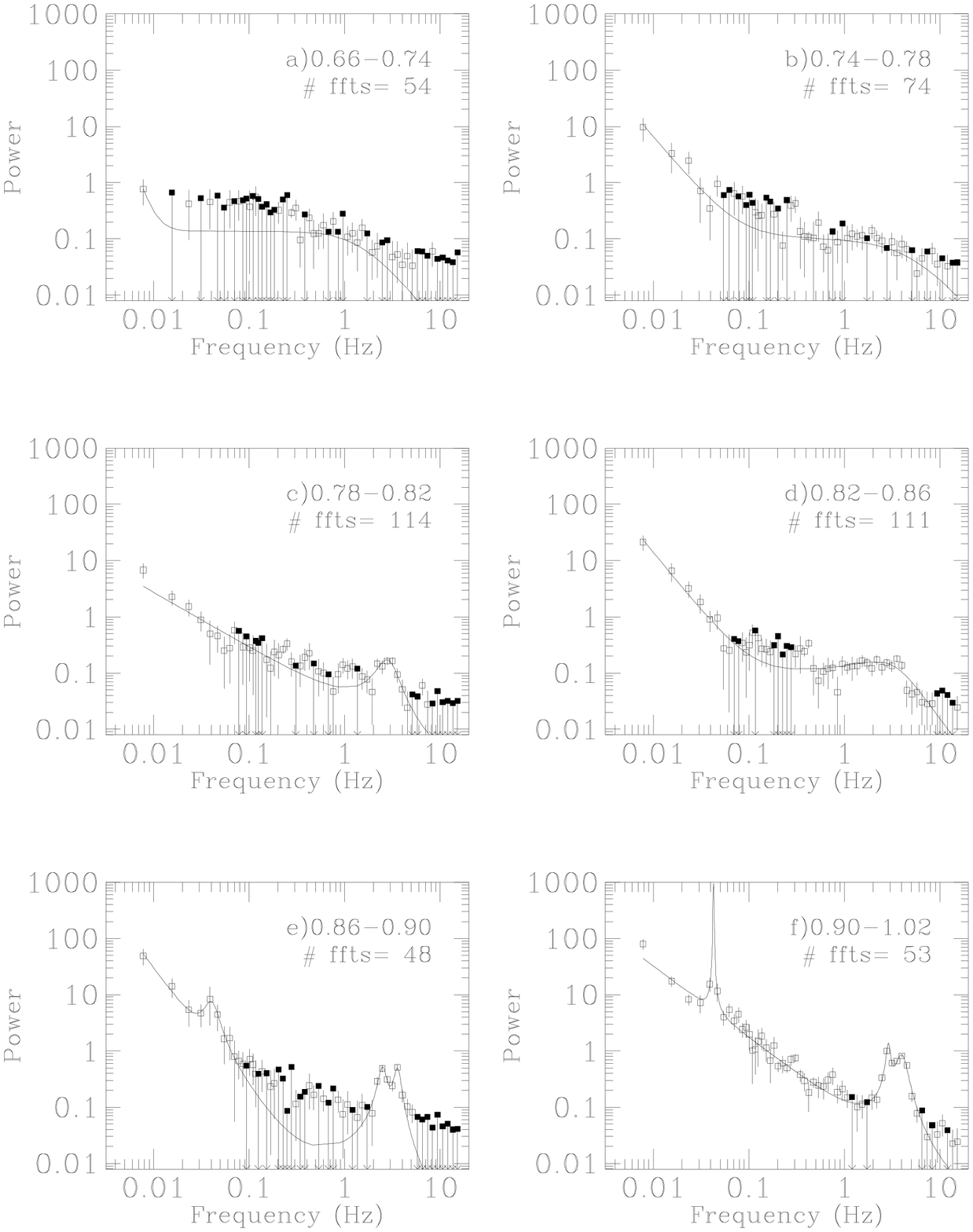 hoffset=-60 voffset=-56}{14.7cm}{21.5cm}
\FigNum{\ref{fig:pepds}}
\end{figure}

\clearpage
\pagestyle{empty}
\begin{figure}
\PSbox{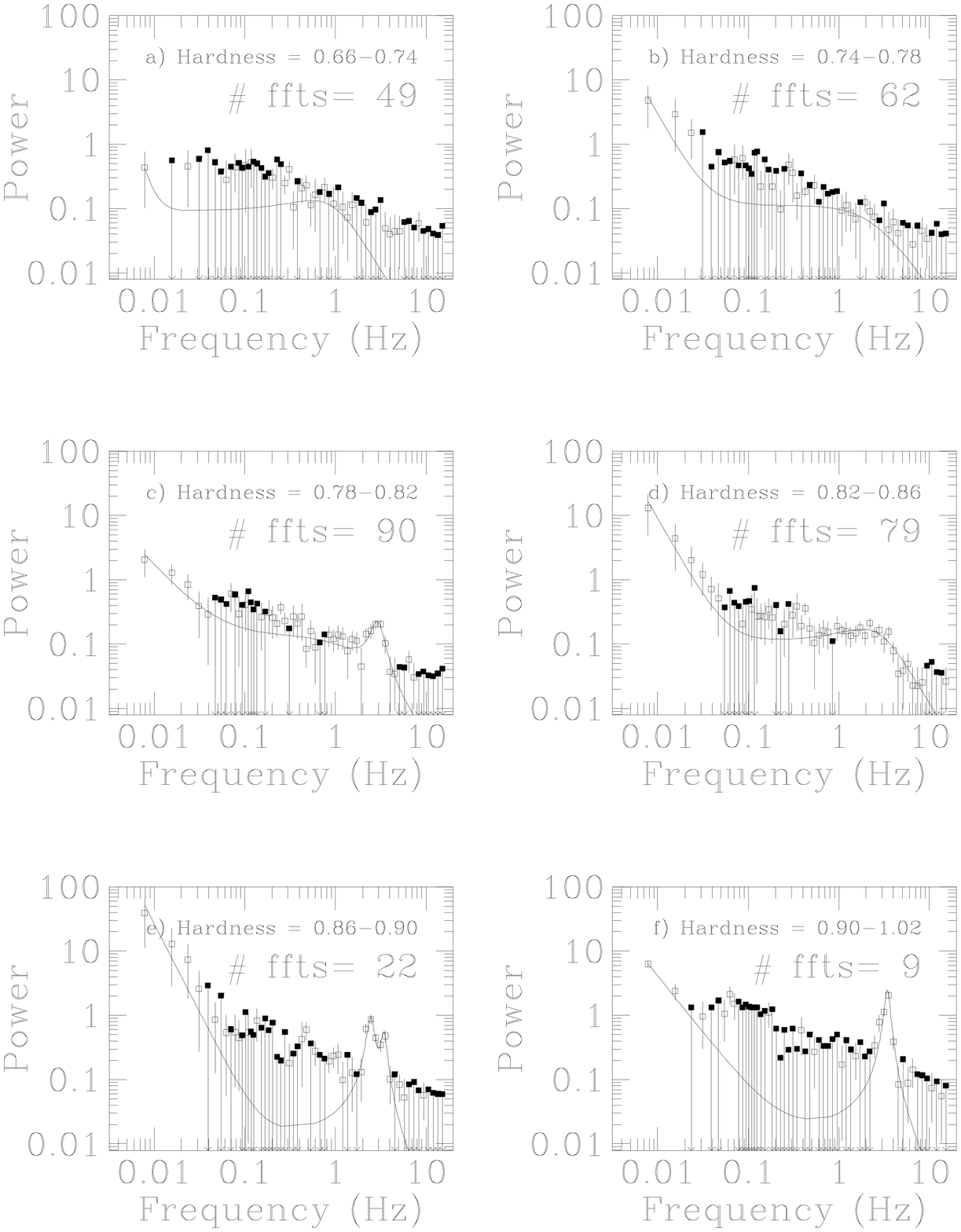 hoffset=-60 voffset=-56}{14.7cm}{21.5cm}
\FigNum{\ref{fig:nohumppds}}
\end{figure}

\clearpage
\pagestyle{empty}
\begin{figure}
\PSbox{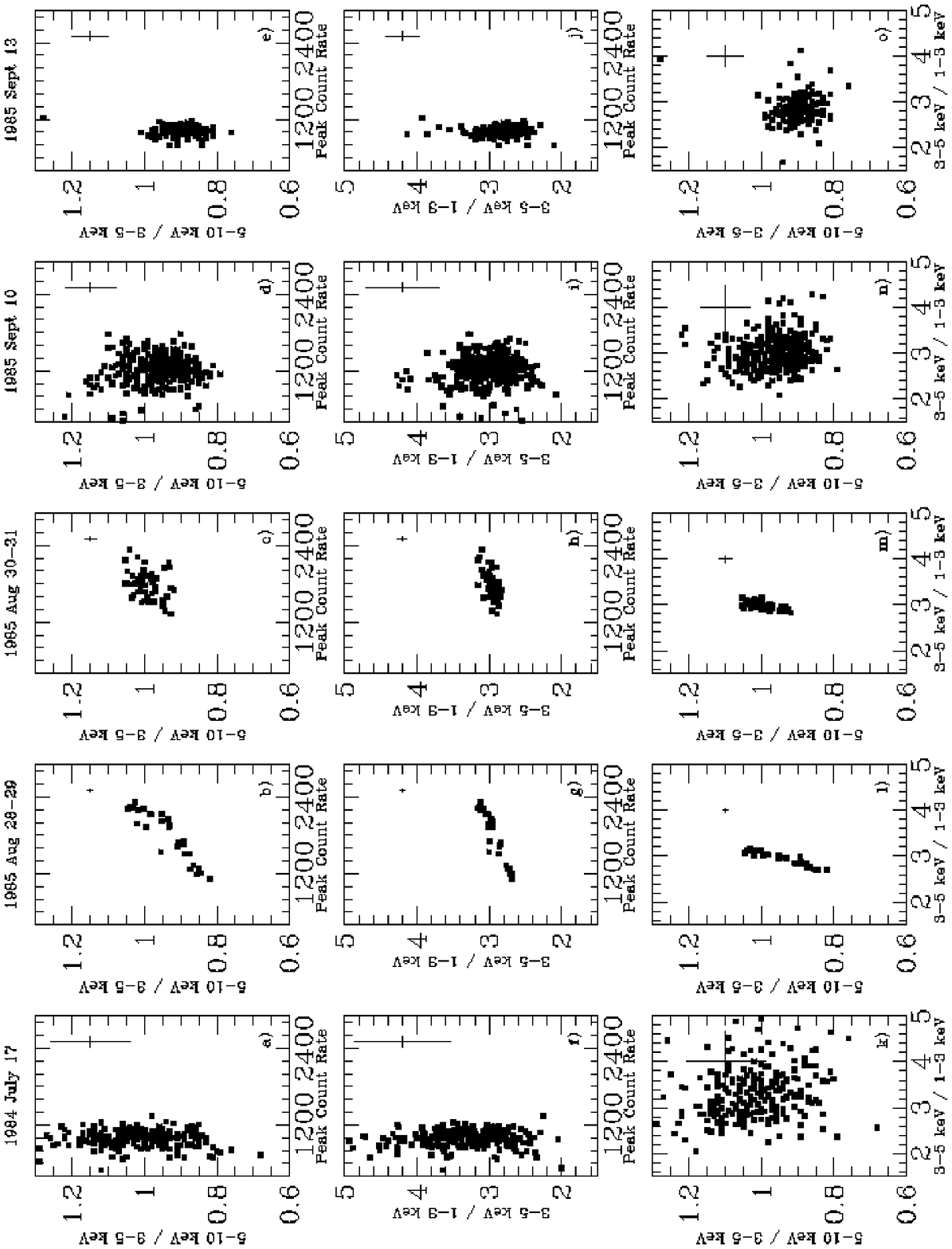 hoffset=-80 voffset=-56}{14.7cm}{21.5cm}
\FigNum{\ref{fig:hidccdcomp}}
\end{figure}

\clearpage
\pagestyle{empty}
\begin{figure}
\PSbox{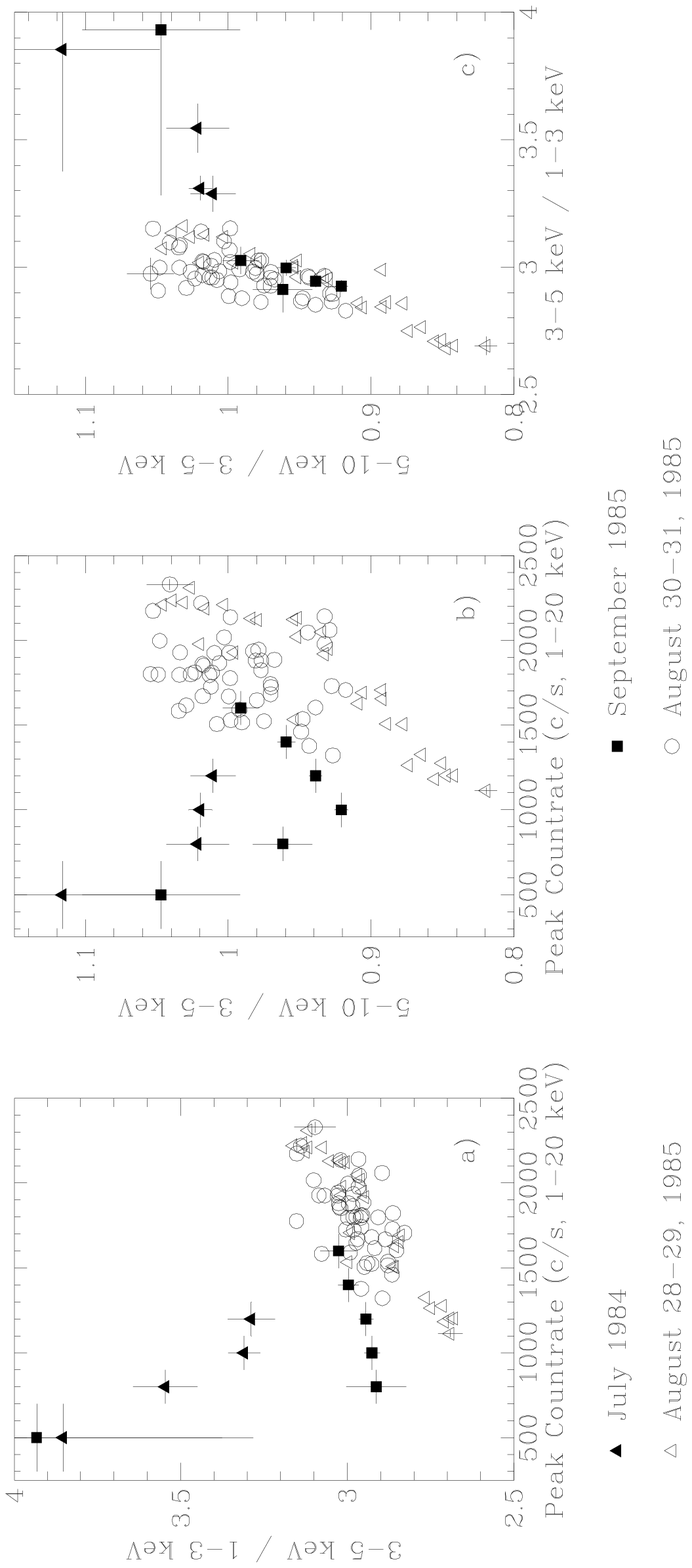 hoffset=-80 voffset=-56}{14.7cm}{21.5cm}
\FigNum{\ref{fig:burstcomp}}
\end{figure}

\clearpage
\pagestyle{empty}
\begin{figure}
\PSbox{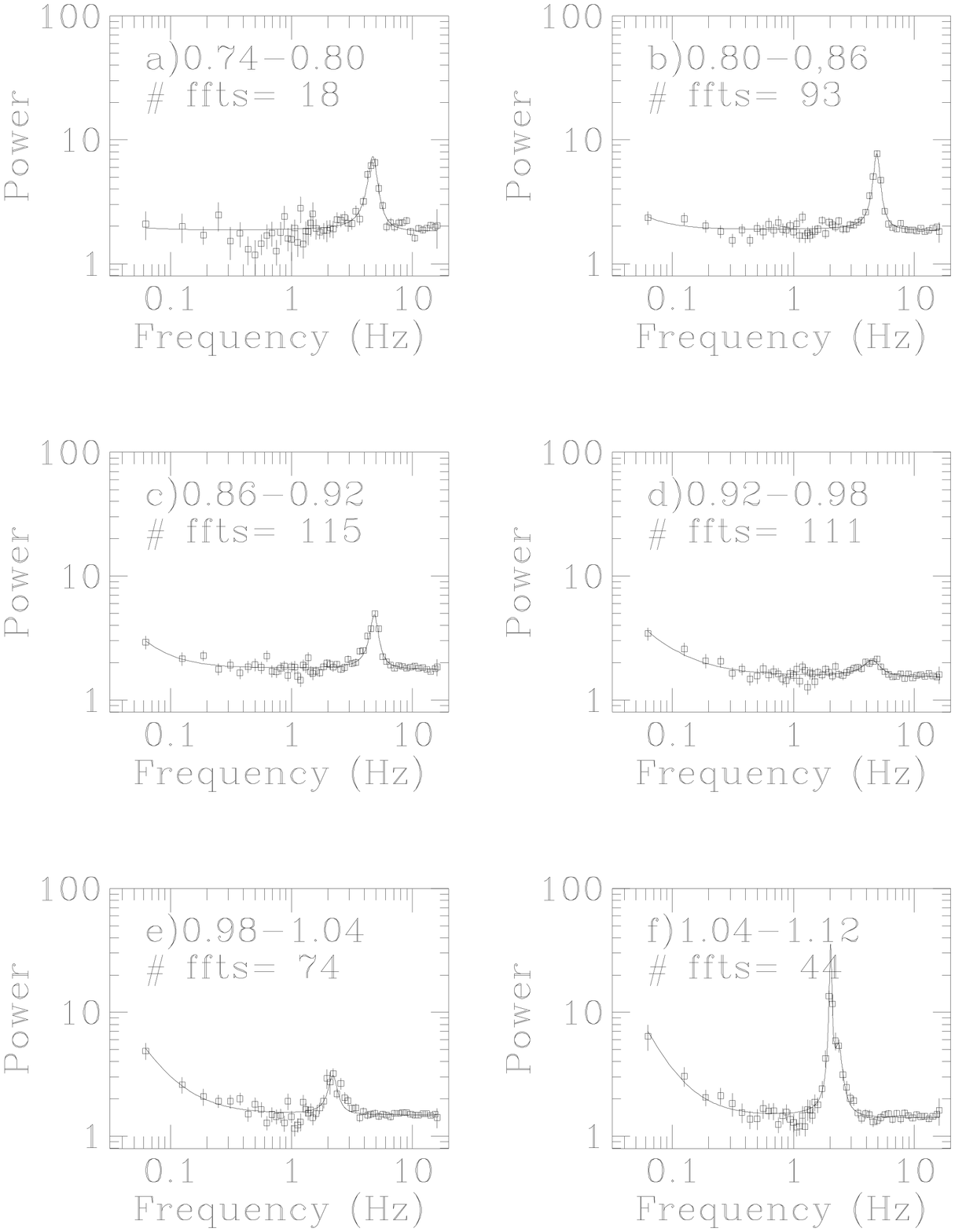 hoffset=-80 voffset=-56}{14.7cm}{21.5cm}
\FigNum{\ref{fig:burstpds}}
\end{figure}

\clearpage
\pagestyle{empty}
\begin{figure}
\PSbox{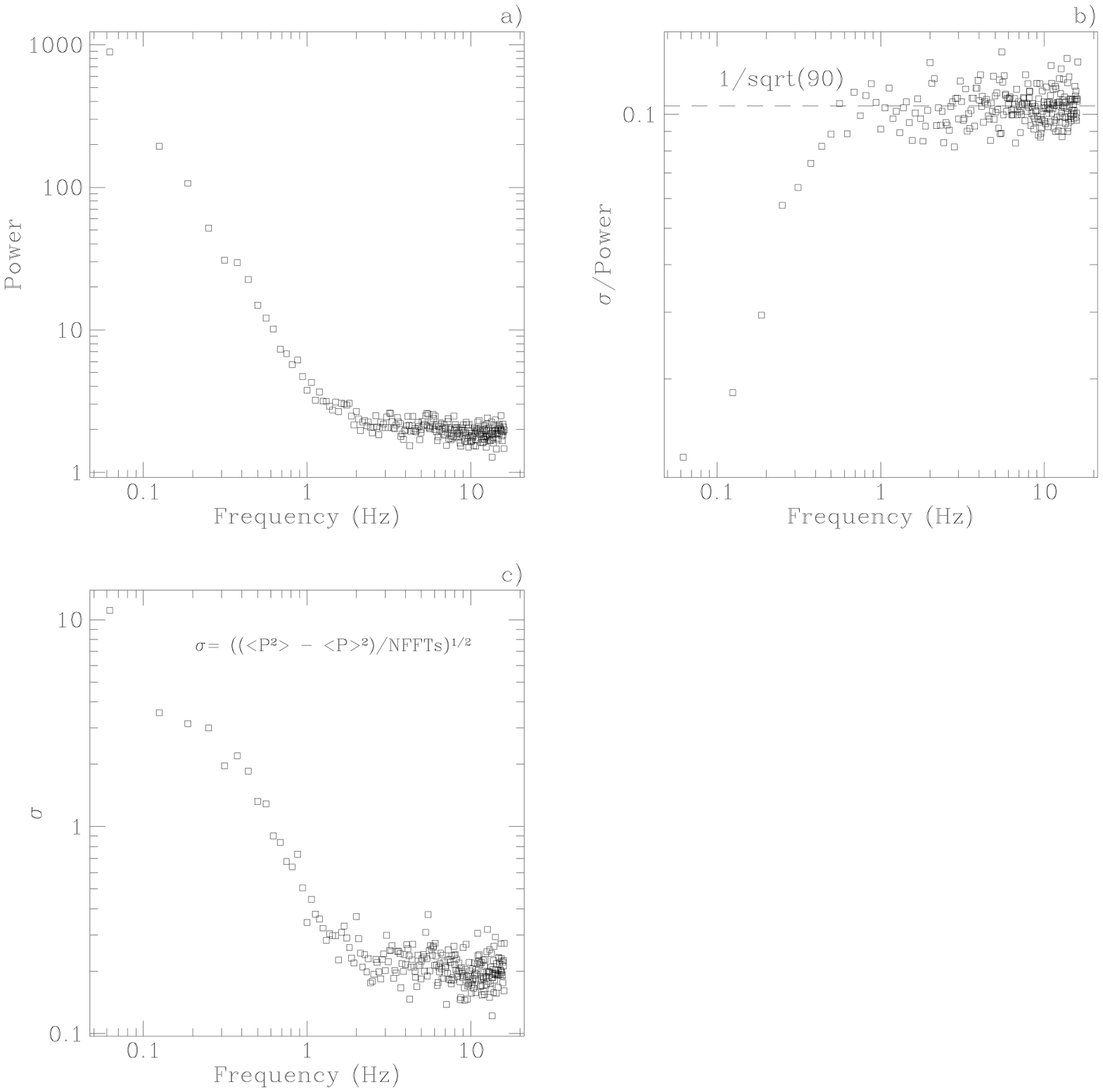 hoffset=-80 voffset=-56}{14.7cm}{21.5cm}
\FigNum{\ref{fig:notcesme}}
\end{figure}

\clearpage
\pagestyle{empty}
\begin{figure}
\PSbox{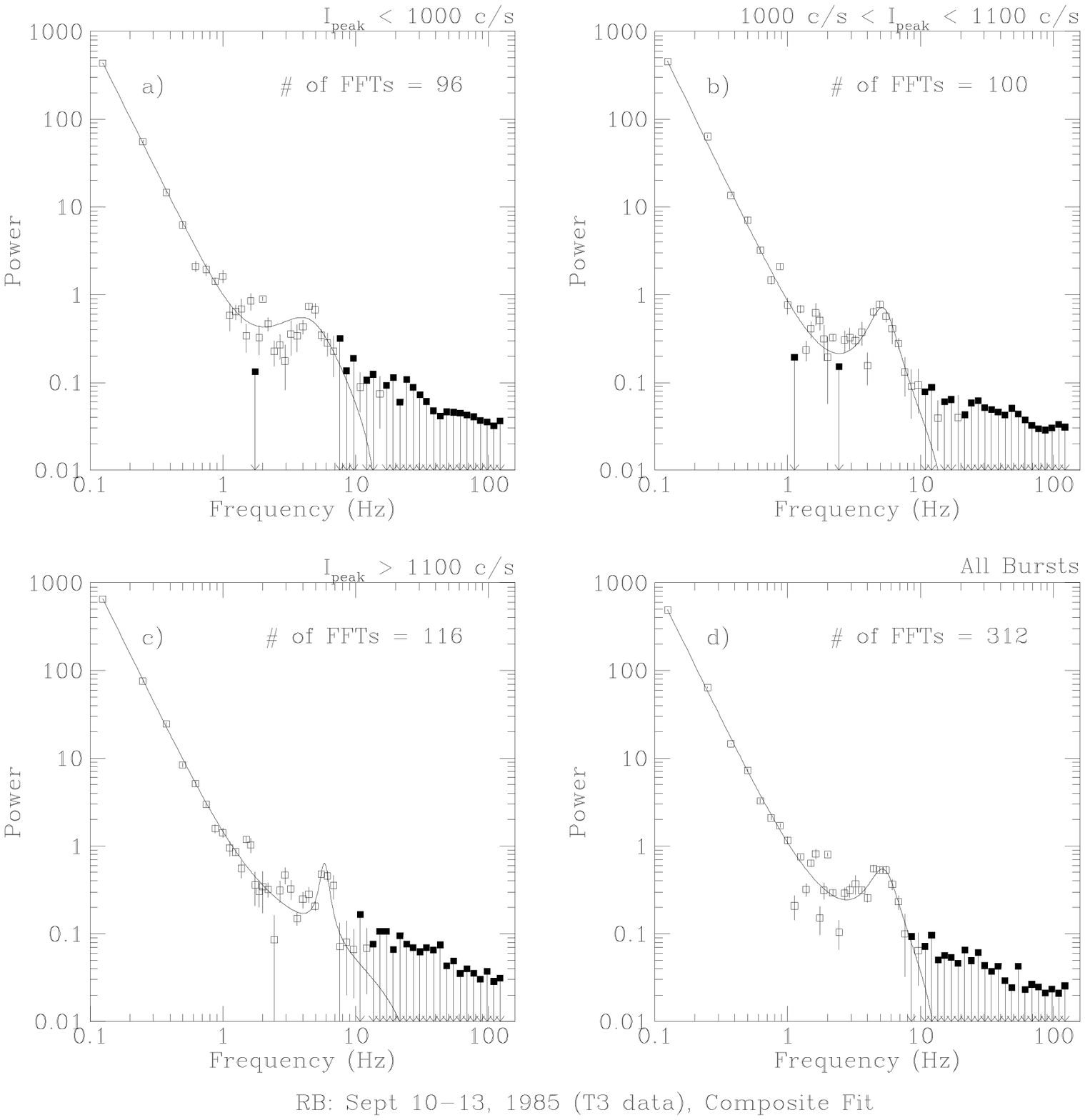 hoffset=-80 voffset=-56}{14.7cm}{21.5cm}
\FigNum{\ref{fig:985totpds} }
\end{figure}

\clearpage
\pagestyle{empty}
\begin{figure}
\PSbox{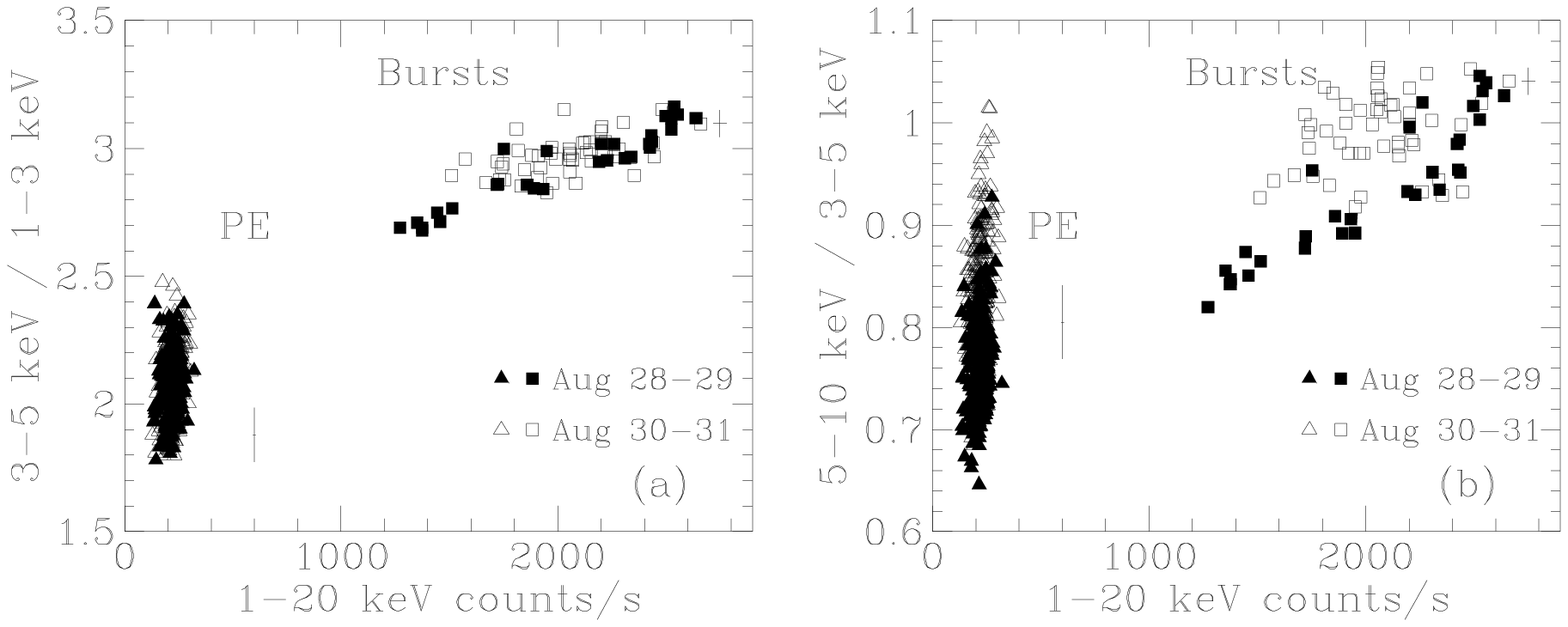 hoffset=-80 voffset=-56}{14.7cm}{21.5cm}
\FigNum{\ref{fig:HID85}}
\end{figure}

\clearpage
\pagestyle{empty}
\begin{figure}
\PSbox{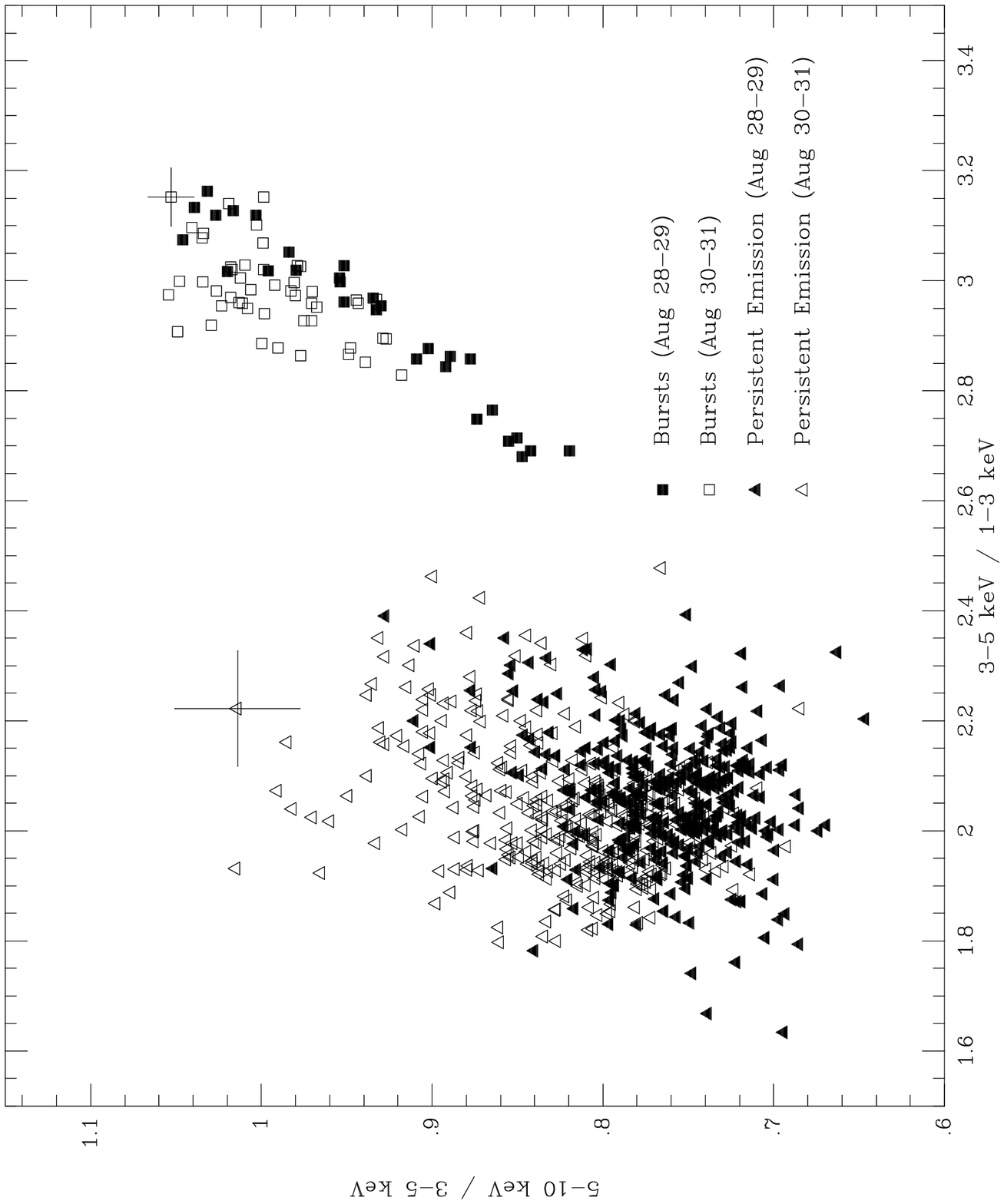 hoffset=-80 voffset=-56}{14.7cm}{21.5cm}
\FigNum{\ref{fig:85CCDcomp}}
\end{figure}

\clearpage
\pagestyle{empty}
\begin{figure}
\PSbox{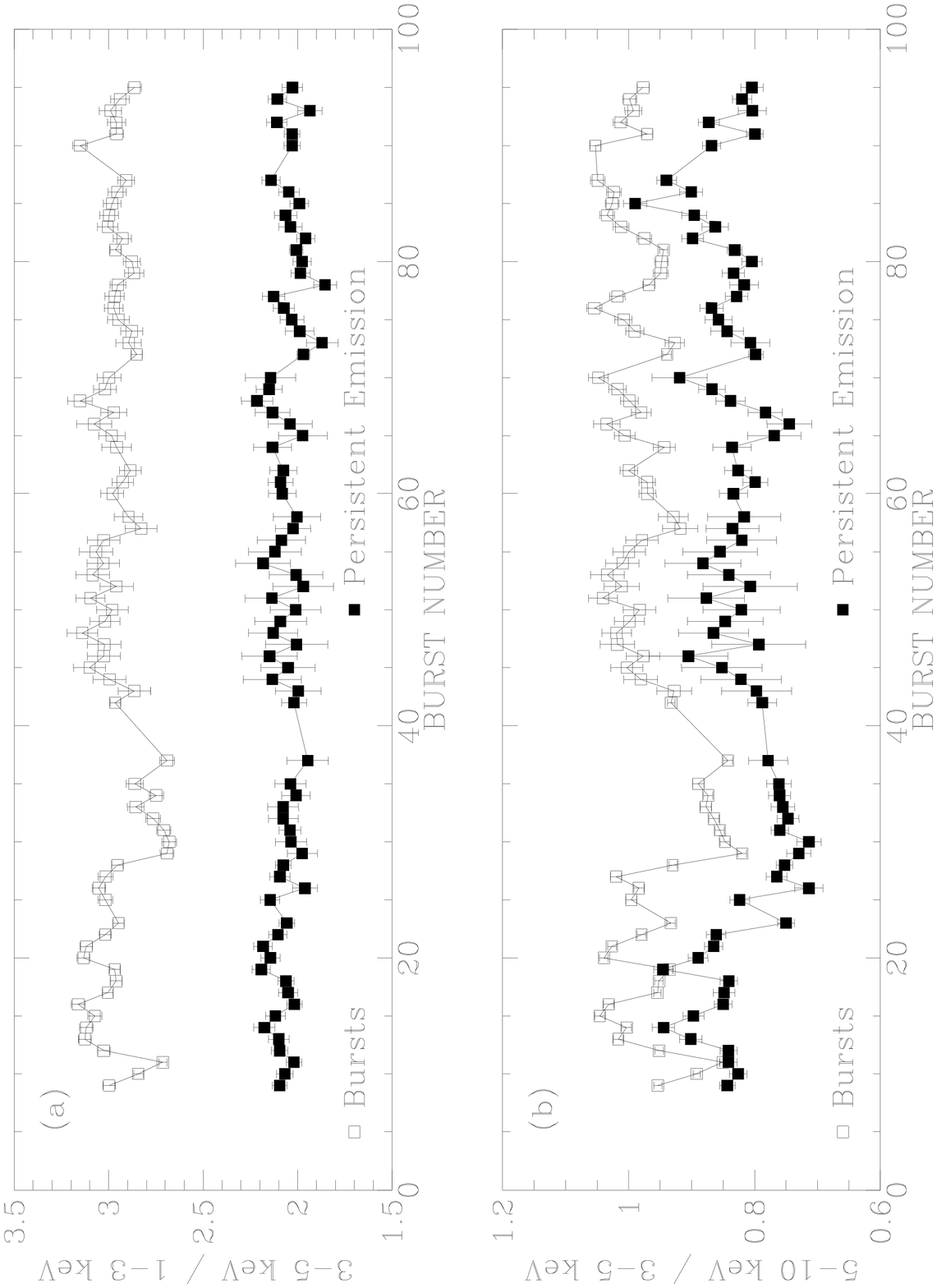 hoffset=-80 voffset=-56}{14.7cm}{21.5cm}
\FigNum{\ref{fig:burstpe}}
\end{figure}

\clearpage
\pagestyle{empty}
\begin{figure}
\PSbox{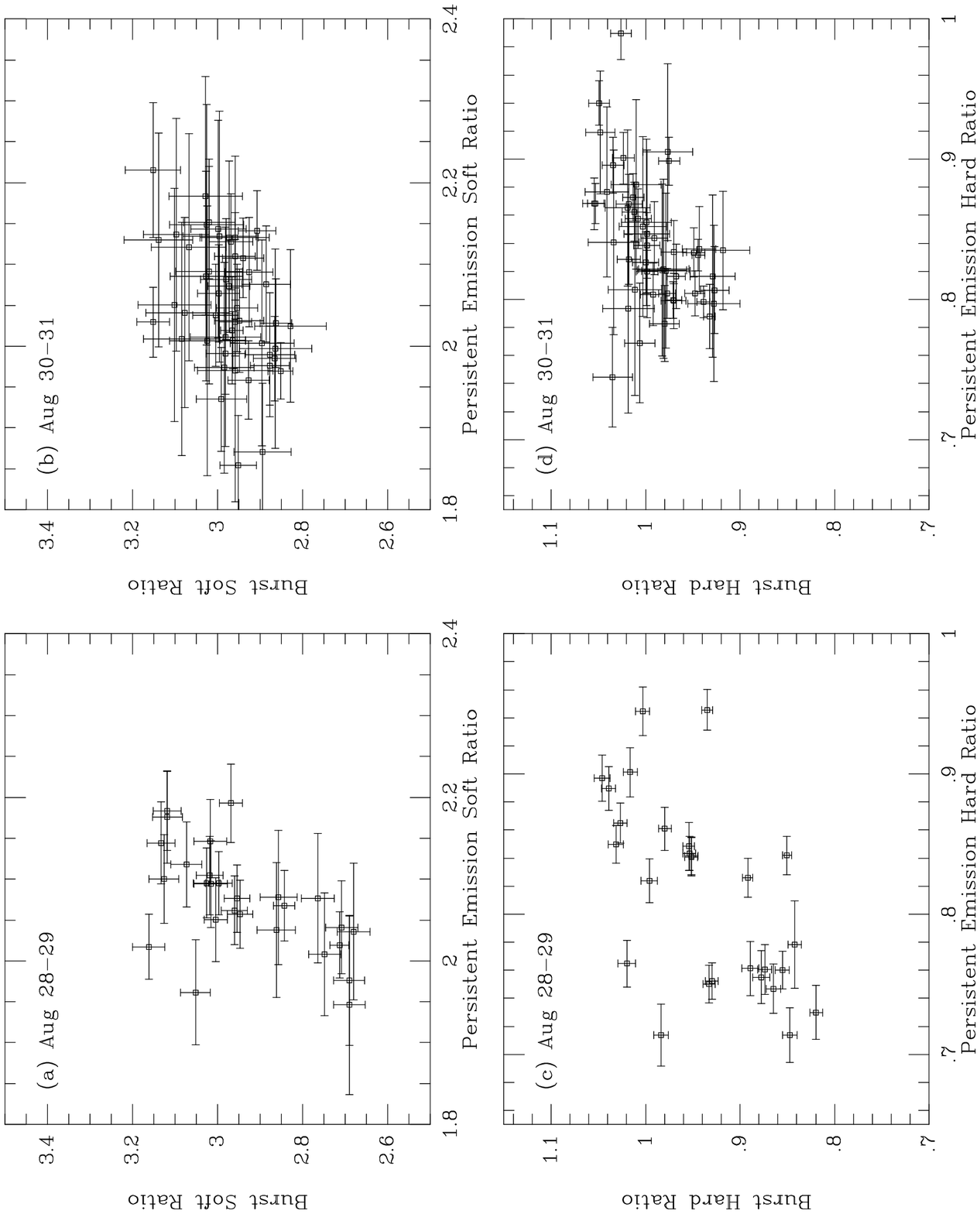 hoffset=-80 voffset=-56}{14.7cm}{21.5cm}
\FigNum{\ref{fig:hardbpe}}
\end{figure}

\clearpage
\pagestyle{empty}
\begin{figure}
\PSbox{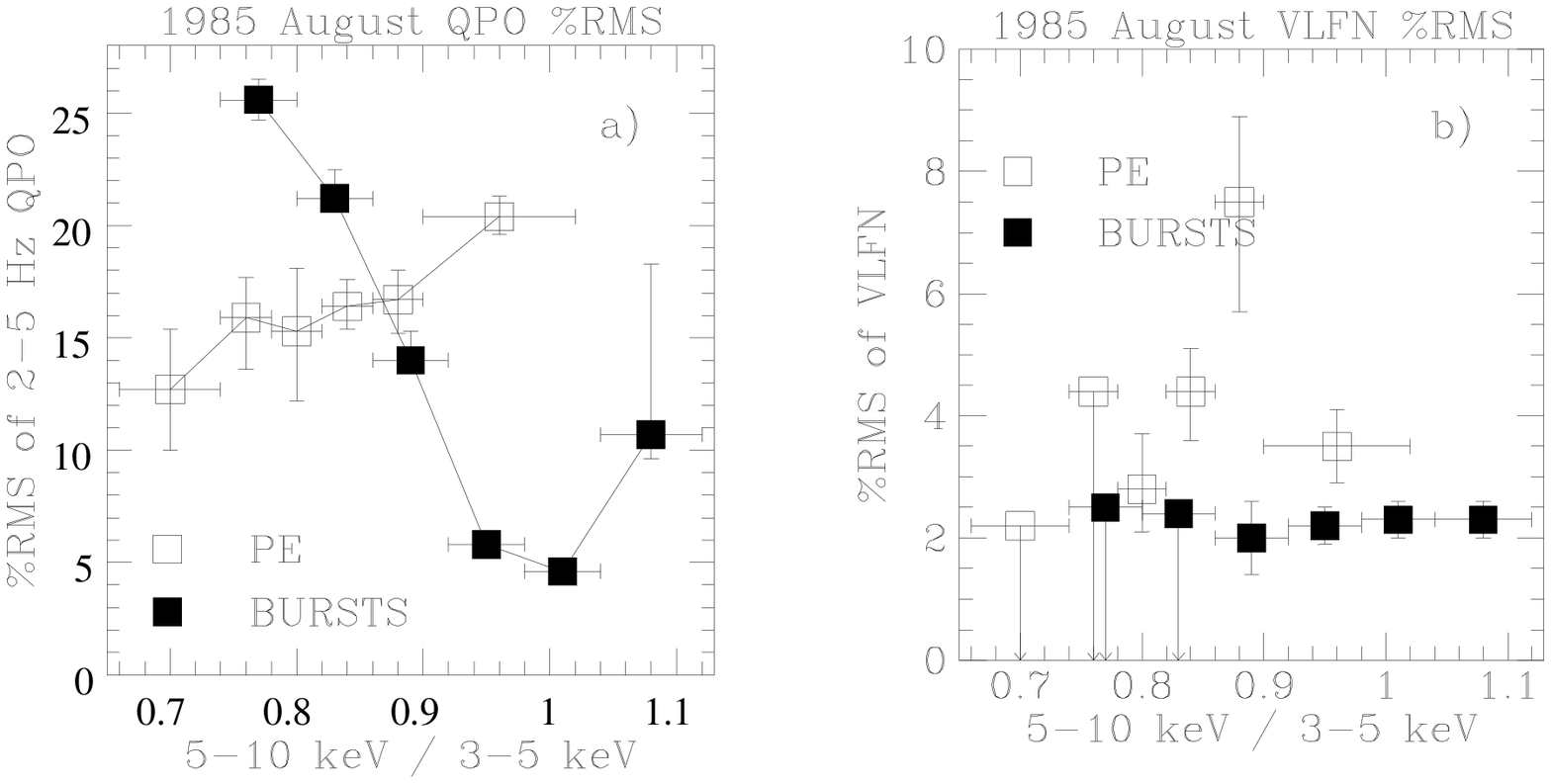 hoffset=-80 voffset=-56}{14.7cm}{21.5cm}
\FigNum{\ref{fig:bupermscomp}}
\end{figure}

\clearpage
\pagestyle{empty}
\begin{figure}
\PSbox{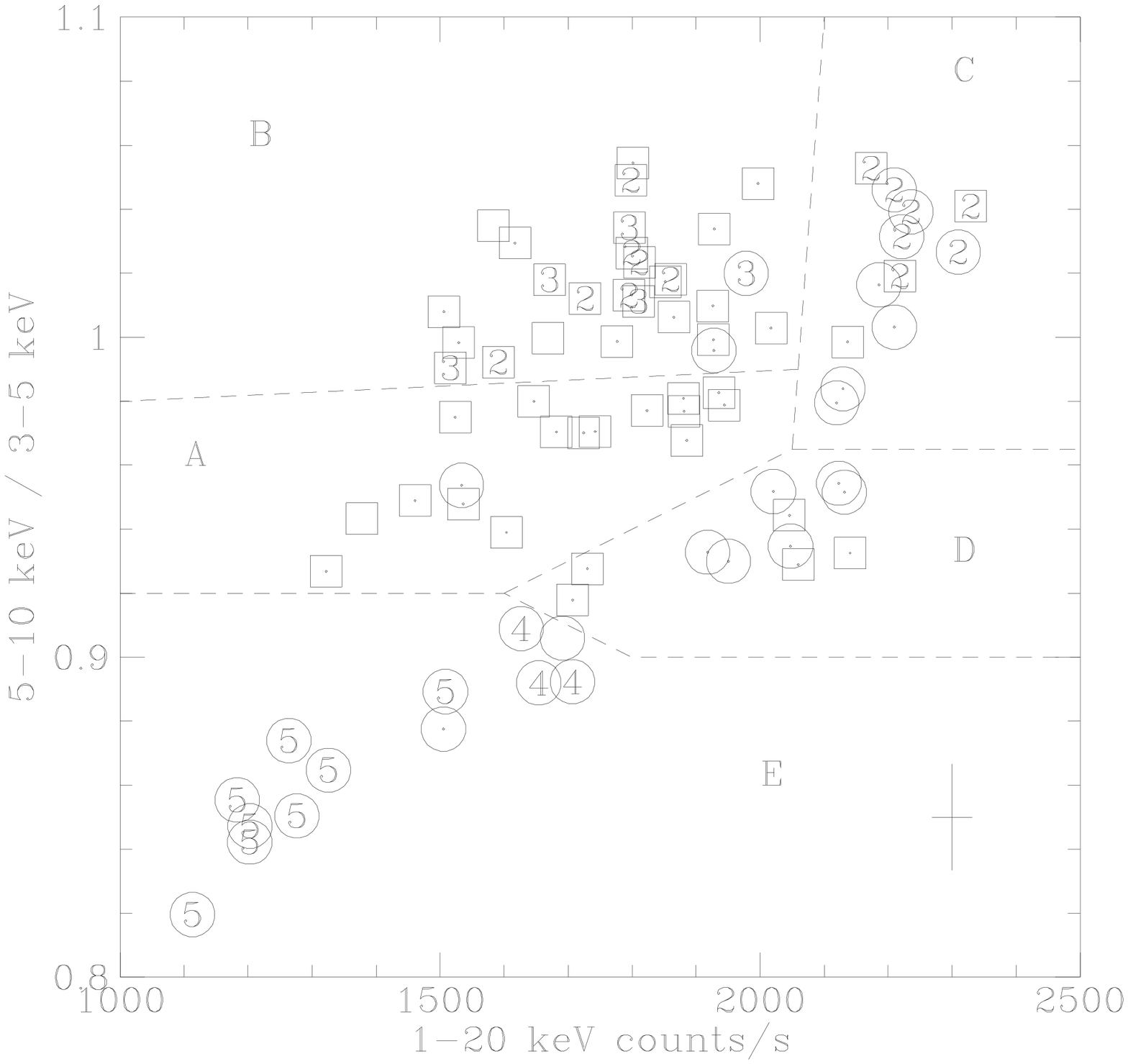 hoffset=-50 voffset=-56}{14.7cm}{21.5cm}
\FigNum{\ref{fig:z}}
\end{figure}

\clearpage
\pagestyle{empty}
\begin{figure}
\PSbox{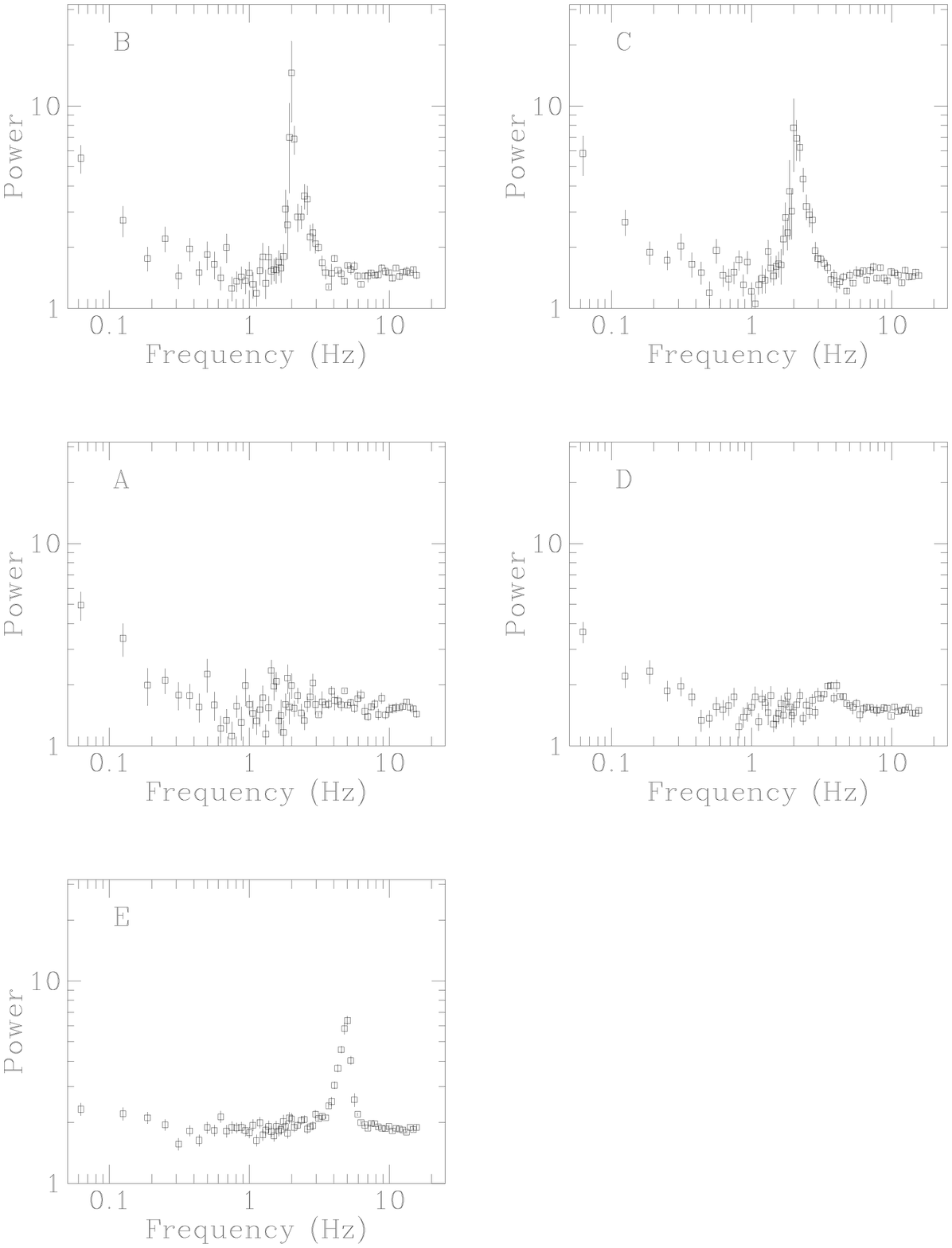 hoffset=-80 voffset=-56}{14.7cm}{21.5cm}
\FigNum{\ref{fig:zpds}}
\end{figure}

\end{document}